\documentstyle[11pt]{article}

\catcode`@=11
\@addtoreset{equation}{section}
\catcode`@=12

\def\alt{{\lower2mm\hbox{$\,\stackrel{\textstyle <}{\sim}\, $}}}

\begin{document}

\begin{titlepage}
\setcounter{page}{1}

\title{\small\centerline{July 1993 \hfill DOE-ER\,40757-022}
\small\rightline{(Slightly modified November 1993) \hfill CPP-93-22}\medskip
{\LARGE\bf Proton decay and related processes\\[3mm]
in unified models with gauged baryon number}\bigskip\bigskip}

\author{Palash B. Pal\\
\normalsize \em Center for Particle Physics, University of Texas,
              Austin, TX 78712, USA
\medskip\\
Utpal Sarkar\\
\normalsize \em Theory Group, Physical Research Laboratory,
Ahmedabad-380009, INDIA}

\date{}
\maketitle
\rm
\vfill
\begin{abstract} \normalsize\noindent
In unification models based on SU(15) or SU(16), baryon number is part
of the gauge symmetry, broken spontaneously. In such models,
 we discuss various
scenarios of important baryon number violating processes like proton
decay and neutron-antineutron oscillation. Our analysis depends on
the effective operator method, and covers many variations of symmetry
breaking, including different intermediate groups and different Higgs
boson content. We discuss processes mediated by gauge bosons and Higgs
bosons parallely.  We show how accidental global or
discrete symmetries present in the full gauge invariant Lagrangian
restrict baryon number violating processes in these models. In all
cases, we find that baryon number violating interactions are
sufficiently suppressed to allow grand unification at energies much
lower than the usual $10^{16}$ GeV.
\end{abstract}
\vfill

\thispagestyle{empty}
\end{titlepage}
%

\section{Introduction}
                 In all gauge theories reasonably verified by
experiments, fermions transform as the fundamental representations of
the non-abelian gauge groups.  Quarks
transform as the fundamental representation of the color group SU(3),
left-handed fermions are fundamental representations of the
electroweak SU(2). Inspired by this, it is  intriguing to consider the
idea that all
fermions transform like the fundamental representation of the grand
unified gauge group. This leads to grand unified models based on the
maximal symmetry group \cite{PSS75} for each generation,  SU(16),
where the fermions all appear in the fundamental multiplet:
	\begin{eqnarray}
\Psi_L \equiv \left( u_r u_b u_y \, d_r d_b d_y \;
\widehat d_r \widehat d_b \widehat d_y \, \widehat u_r \widehat u_b
\widehat u_y \;  \nu_e e^- e^+ \widehat\nu_e \right)_L \,.
\label{fermions}
	\end{eqnarray}
The indices $r,b,y$ are three colors, and hats denote antiparticles
 for any fermion field $\psi$:
	\begin{eqnarray}
\widehat \psi = C \gamma^0 \psi^* \,,
	\end{eqnarray}
where $C\gamma_\mu C^{-1}=-\gamma_\mu^T$.
Thus, for example, $\widehat\nu_{eL}$ is the antiparticle of the right
handed neutrino $\nu_{eR}$, assuming that it exists. The same
pattern is repeated for other generations. Mirror fermions are needed
to cancel the anomalies. One important feature of this model is that
both baryon number ($B$) and lepton number ($L$), which are known
symmetries of low energy physics, appear as gauge symmetries at high
energy. In fact, this was one of the main motivations of Pati, Salam
and Strathdee who first introduced such models \cite{PSS75}.

A new variant of these models has received some attention lately, where
the gauge group is SU(15) \cite{Adler89,FrLe90}. The difference with
SU(16) is that the right handed neutrinos, which are not confirmed
experimentally, are assumed not to exist, so that there are only 15
left-chiral fermionic fields per generation. Baryon number is still
part of the gauge symmetry although lepton number is not.

The interesting point about these models is that their characteristics
are very different from the standard unification models based on the
gauge groups SU(5), SO(10), $E_6$ etc. For example, it has been shown
that renormalization group analysis of certain symmetry breaking
chains of these models yield low unification scales
\cite{Adler89,FrLe90,Pal,BSMS92}, as low as $10^8$ GeV in some cases.
All the known chains with such low unification scale have the property
that they all break the unified group in such a way that at
intermediate scales, quarks and leptons transform under separate
subgroups of the gauge group.  Because of low unification scales,
these models do not suffer from the cosmological monopole problem
\cite{Pal91}, in sharp difference with SU(5) models. Important and
interesting constraints on rare processes can be put in these models
\cite{Pal,Pal93}. Although many of these points were first made with
the SU(15) gauge group, it is now known that there are symmetry
breaking chains of the original SU(16) gauge group as well which show
these characteristics~\cite{DKP93,Bra93,Lav94}.

One crucial question  arises now. How can a low unification scale be
consistent with known bounds on proton lifetime?  Of course,
it is easy to see that gauge interactions do not violate $B$ in the
unbroken phase. This is another important
difference with SU(5), SO(10) or $E_6$ models. With a limited number
of Higgs multiplets, it was argued that baryon number
symmetry ($B$) is not violated \cite{FrLe90} even after symmetry
breaking has taken place.  Subsequently, it was emphasized
\cite{Pal,SMS} that since $B$ is part of the gauge symmetry of these
models, it must be broken spontaneously in order to avoid a massless
gauge boson corresponding to an unbroken $B$ symmetry,  and therefore
the Higgs sector must be expanded.

Once this was pointed out, various scenarios of proton decay were
considered by different authors \cite{FrKe90,Pal92,BSMS92,DKP93}.
Particularly powerful is the method of effective operators, which will
be explained in Sec.~\ref{s:general}.  A simple dimensional analysis
performed with these effective operators \cite{Pal92,DKP93} shows that
proton decay amplitude in these models are suppressed by as many as
the fifth power of the grand unification mass, as opposed to the
second power in the case of standard unification models. Because of
this reason, low unification scale is consistent with the known
bounds on the proton lifetime.

However, this analysis was performed only with proton decay mediated
by Higgs bosons, in a handful of scenarios. But there is another kind
of contribution to the proton decay amplitude in these models.
It is true that in the unbroken theory, each gauge boson carries a
well-defined baryon number and therefore cannot mediate $B$-violating
processes \cite{Adler89,FrLe90}. However, once the gauge group is
spontaneously broken, gauge bosons with different baryon numbers can
mix with one another and therefore the mass eigenstates of gauge
bosons are not, in general, eigenstates of baryon number. They can
therefore mediate baryon number violating processes. Although such
contributions were discussed in some detail \cite{MoPo82,RaSa82}
in some early papers on the SU(16) model, the possibility of low
unification scale was not realized at that time. In the context of low
energy unification, it was discussed briefly only at the tree level in
a very specific scenario~\cite{BSMS92}.

In the present paper, our focus is threefold. First, we discuss not
only proton decay which is a $|\delta B|=1$ process, but also
neutron-antineutron oscillation which is a $|\delta B|=2$ process to
see whether both are consistent with low energy unification. Of
course, even when we will be explicitly talking
about ``proton decay'', the comments can easily be translated to the
baryon number violating decays of the neutron as well. Second,
we include both gauge boson mediated and Higgs boson mediated
processes in our analysis.  Third, we point out accidental global
symmetries in the full gauge invariant Lagrangians of the models which
seriously constrain possible baryon number violating processes. For
example, in one case we find that there are no operators involving
four fermions which can give baryon number violation even after baryon
number symmetry is spontaneously violated. In such cases, we extend
the analysis of proton decay to operators involving six fermions,
which has not been done before.

\section{General considerations}\label{s:general}
                        From the symmetries of the standard model
alone, one can argue that the dimension of any proton
decay operators must be six or higher \cite{Wei79,WiZe79}, whereas for
$|\Delta B|=2$ processes it is at least nine \cite{Wei80}. Thus, we
should be looking at non-renormalizable operators generated by the
theory. In general, these operators can involve both ordinary and
mirror fermions since the physical up quark, for example, can be a
superposition of the ordinary and the mirror quark fields. However,
for the sake of simplicity, we will assume that the mirrors do not mix
with ordinary fermions. This can be attained naturally if we impose a
discrete symmetry, $\Psi^{(M)} \to -\Psi^{(M)}$ where $\Psi^{(M)}$
stands for the mirror fields. This is the most popular discrete
symmetry considered in the context of mirror fermions. In the presence
of this symmetry the model ceases to remain vectorial and hence the
fermion masses are protected. Hence the survival hypothesis is
applicable to these theories. The mirrors can be heavier than the
ordinary fermions if their Yukawa couplings are consistently larger.
In that case, they will not figure in any of the low energy processes
we will be discussing. We will also assume that there are no hitherto
unknown bosons lighter than the nucleon mass. Low energy operators
involving nucleons should then involve ordinary fermionic fields only.

To analyze these operators, we adopt the procedure used in Refs.\
\cite{DPY91,Pal92}, where one first constructs effective operators
which are invariant under the gauge group.    For baryon number
violating processes at low energy, the full gauge invariant operators
will in general also contain some scalar fields so that, when these
scalars develop vacuum expectation values (VEVs), one obtains
operators involving the fermionic fields only. If the VEVs are baryon
number violating, the fermion field operator generated after putting
the VEVs would violate baryon number. This is the main difference with
operator analysis for proton decay performed in the context of SU(5)
or SO(10) grand-unified models \cite{Wei79,WiZe79}, where the
operators have to obey only the symmetries of the standard model since
the unbroken grand unified model does not conserve baryon number.
Here, the unbroken operators must obey the symmetries of the unified
model, which of course is much larger than that of the standard model.
This requirement severely restricts the type of baryon number
violating operators that one can construct.

                          The above discussion is applicable equally
for any baryon number violating process induced by Higgs-boson
and gauge-boson exchanges.  We now consider proton decay
in particular. Here, one needs an operator where the number of
fermionic fields is at least four \cite{Wei79}, and in most part we
will discuss operators where the number is in fact four, except in
Sec.~\ref{s:154} where we will find that such operators are forbidden
for proton decay.

  For Higgs-boson mediated processes, the relevant
bilinears are $(\Psi_L)^TC\Psi_L$, where $C$ is the conjugation matrix
for fermions. For gauge-boson mediated processes, the relevant
bilinears are $\overline{\Psi}_L \gamma_\lambda \Psi_L$. The important
difference is that the first $\Psi$ appears with a complex conjugation
here. If we put in the gauge indices, this will mean a lower index for
the first $\Psi$ and an upper index for the second one. Thus, the
gauge invariants constructed will have a quite different nature than
the ones for the Higgs-boson mediated processes.

The discussion so far implies that, for processes mediated by gauge
bosons, the 4-fermionic operators must appear in the effective
operator in the combination
          \begin{eqnarray}
{\cal K}
\left[ (\overline{\Psi_L})_i \gamma^\lambda (\Psi_L)^j \right] \,
\left[ (\overline{\Psi_L})_k \gamma_\lambda (\Psi_L)^l \right]
 \,,
\label{4fgauge}
          \end{eqnarray}
where $i,j,k,l$ denote gauge indices, and we have suppressed the
generation indices. On the other hand, for Higgs boson mediated
processes, the combinations should be
          \begin{eqnarray}
{\cal K}
\left[ ({\Psi_L})^i C (\Psi_L)^j \right] \,
\left[ ({\Psi_L})^k C (\Psi_L)^l \right]  \,.
\label{4fhiggs}
          \end{eqnarray}
One can think of other operators like $\left[ ({\widehat\Psi_R})_i C
(\widehat\Psi_R)_j \right] \,  \left[ ({\widehat\Psi_R})_k C
(\widehat\Psi_R)_l \right]$ or $\left[ ({\Psi_L})^i C (\Psi_L)^j
\right] \,  \left[ ({\widehat\Psi_R})_k C (\widehat\Psi_R)_l \right]$,
where $\widehat\Psi_R$ is the multiplet which contains the
antiparticles of the fields in $\Psi_L$, but these are
either just hermitian conjugates of the operators in Eqs.\
(\ref{4fgauge}) and (\ref{4fhiggs}), or can be Fierz transformed to
them. So, we need not discuss them separately.

 Our goal is to find gauge invariant operators which can
give rise to the four-fermion operators of Eqs.\  (\ref{4fgauge}) and
(\ref{4fhiggs}) after symmetry breaking.  This discussion
involves the Higgs content
of the model and the precise way in which baryon number is violated,
and therefore has to be done separately for SU(15) and SU(16). This
will be done in the ensuing sections. Also, as we said before, we will
encounter specific models where 4-fermion operators are inconsistent
with the symmetries of the model. For such cases, we describe here the
general case where the fermionic part of the operator has $2n$ number
of fields. Let us denote such an operator symbolically as ${\cal
K}_{(2n)} \psi^{2n}$.  The co-efficient ${\cal K}_{(2n)}$ has a mass
dimension $4-3n$. Thus, neglecting the masses of all decay products, a
simple dimensional analysis will give
           \begin{eqnarray}
\tau_p^{-1} \approx m_p^{6n-7} \, {\cal K}_{(2n)}^2 \,.
           \end{eqnarray}
Then, the known limits on the proton lifetime,\footnote{The precise
bounds depend on the specific decay mode. For rough estimates, we use
the same bound for all modes.}
           \begin{eqnarray}
\tau_p > 3 \times 10^{32} \, {\rm yr},
           \end{eqnarray}
implies
           \begin{eqnarray}
{\cal K}_{(2n)} \alt 10^{-32} \, m_p^{4-3n}.
\label{K2nbound}
           \end{eqnarray}
In the specific case where $n=2$ (which is the most frequent case so
that we will omit the subscript of ${\cal K}$ in this case), we obtain
           \begin{eqnarray}
{\cal K} \alt 10^{-32} \, {\rm GeV}^{-2}.
\label{Kbound}
           \end{eqnarray}
In standard unification models like SU(5) or SO(10), ${\cal K}\simeq
g^2 M_G^{-2}$, so that one needs $M_G/g>10^{16}$ GeV in
order for the models
to be phenomenologically viable. In the models that we consider, we
will see that ${\cal K}$ is further suppressed by ratios of different
mass scales, and that is why smaller unification scales will be
consistent with phenomenology.

\section{Scenarios of baryon number violation in SU(15)
models}\label{s:15}
         Various scenarios of baryon number violation has been
discussed in the literature \cite{FrKe90,Pal92,BSMS92} in the context
of the SU(15) gauge group. We present some such chains later in
this section. In all these
chains, for all symmery breakings above the weak scale, we use Higgs
bosons either in completely antisymmetric representations, or in the
adjoint representation or representations which can be obtained by
taking tensor products of two adjoint representations. This is done
for the sake of economy and definiteness. We also assume that, unless
mentioned otherwise, the only Higgs boson multiplets present in the
model are the ones which have VEVs.

At the weak scale, however, we make an exception. Here, unless
otherwise specified, we assume that the symmetry breaking is performed
by a {\em symmetric} rank-2 multiplet $S$. This is motivated
phenomenologically.  If we use antisymmetric tensor to be the only
field to couple to fermions, the fermion mass matrices would be
antisymmetric. For  three generations, this will imply that
one mass eigenvalue is zero and the other two equal
for particles of any given charge. This is very unrealistic, so we
will not consider this possibility further.

\subsection{Baryon number violated by an antisymmetric rank-3
multiplet}\label{s:153}
\subsubsection{Symmetries of the model}\label{s:153symm}
                    For the SU(15) model, Pal \cite{Pal92}
introduced the most economic Higgs boson spectrum that
leads to breaking pattern with ``un-unified'' intermediate stages. The
multiplets necessary for this purpose  are the
antisymmetric rank-3 multiplet $\Phi^{[ijk]}$, the adjoint $T^i{}_j$,
the symmetric rank-2 multiplet $S^{\{ij\}}$ which gives fermion
masses, and an additional one, $H^{[ij]}_{[kl]}$, which will be called
the antisymmetric bi-adjoint since it appears in the tensor product of
two adjoint representations and both the upper and the lower indices
are antisymmetrized.  Here and henceforth, the square and
curly brackets denote antisymmetrization and symmetrization of
indices.

  In Fig.~\ref{f:153ssb}, we show the complete chain of
symmetry breaking, where the numbers $n$ denote a factor SU($n$) in
the gauge group if $n>1$, and a U(1) factor if $n=1$.  Thus,
at the highest stage, the multiplet $\Phi^{[ijk]}$ develops a VEV
$\left<\Phi^{\nu_ee^-e^+}\right>$, which breaks the unification group
SU(15) down to ${\rm SU(12)}_q\times {\rm SU(3)}_\ell$, where the
subscripts $q$
or $\ell$ denote that only quarks/antiquarks or leptons/antileptons are
non-singlets under the subgroup. This VEV also breaks lepton number by
1 unit.  At the next stage, ${\rm SU(12)}_q$ breaks to ${\rm
SU(6)}_{qL} \times {\rm SU(6)}_{qR} \times {\rm U(1)}_B$. This can be
performed by a VEV in the adjoint representation, as shown in the
figure and explained in the figure caption. At the scale $M_{6qL}$,
the SU(6)$_{qL}$ breaks to its maximal subgroup ${\rm SU(3)}\otimes
{\rm SU(2)}$, under which the fundamental of SU(6) transforms like
(3,2). In SU(6), the lowest dimensional multiplet which has the
component whose VEV can induce this breaking is the 189-dimensional
antisymmetric
bi-adjoint.  Naturally, it is contained in the antisymmetric
bi-adjoint representation of the SU(15).  At the next stage,
SU(6)$_{qR}$ breaks to ${\rm SU(3)}_{uR}\times {\rm SU(3)}_{dR} \times
{\rm U(1)}_{qR\Lambda}$. The SU(3) factors here operate non-trivially
only on the $\widehat u_L$ and $\widehat d_L$ components respectively,
whereas the U(1) quantum numbers are defined to be $+1$ for the
up-type quarks and $-1$ for the down-type ones. Baryon number is
broken at the next stage, where ${\rm SU(3)}_{uR}\times {\rm
SU(3)}_{dR}$ also is broken to the diagonal SU(3) subgroup which is
called SU(3)$_{qR}$. At this stage, the gauge group that appears is
the square of the standard model gauge group, where the quarks and
leptons transform under different SU(2) and different U(1) factors.
This has been discussed under the name ``un-unified'' model by some
authors \cite{ununif}. On the other hand, the right and left chirality
of quarks transform under different color groups, which has been
discussed in the literature under the name ``chiral color''
\cite{RaSa82,chicol}. At the next stage, the standard model gauge group
appears, which is why this scale is called $M_S$.

 It should be understood that some variations of this chain
are obviously possible.  For example,
the scale $M_{6qL}$ can be lower than $M_{6qR}$ or even $M_{3\ell}$.
On the other hand, some scales can merge, so that the standard model
is reached in less number of steps. These will not essentially change
the conclusions of the subsequent discussions and hence will not be
discussed separately. Similar comments apply for other chains which we
will discuss later in the paper.

For this and various other scenarios that we are going to discuss, we
find that the full gauge invariant Lagrangian
involving the specified fields often contains some accidental global
or discrete symmetries  which commute with the gauge
symmetry.  These restrict the type of potentially baryon
number violating operators. To see such symmetries in the present
case, let us write the full Lagrangian in the following suggestive
manner:
	\begin{eqnarray}
{\cal L} = {\cal L}_0 + {\cal L}' \,.
\label{L0L'}
	\end{eqnarray}
Here, ${\cal L}_0$ is the part which is invariant under independent
phase rotations of all complex multiplets present in
the model. These would include, e.g., all gauge interactions,
scalar interactions involving only the adjoint Higgs multiplet and the
antisymmetric bi-adjoint. There will also be some terms involving
other multiplets, e.g., terms like $\Phi^{ijk}\Phi_{ijk}$ or
$S^{ij}S_{jk}S^{kl}S_{li}$.  Obviously, symmetry of ${\cal L}_0$ is
much larger than the gauge symmetry. However, ${\cal L}'$ contains
other terms which are allowed by the gauge symmetry. In the present
case, the Yukawa couplings are the only terms which fall in
this class. Thus,
	\begin{eqnarray}
{\cal L}' = {\cal Y} (\Psi_L)^k (\Psi_L)^l S_{kl} + \mbox{h.c.}\,,
	\end{eqnarray}
where the generational indices on $\Psi_L$ and $\cal Y$ have been
omitted.  However, it is easy
to see that, even with this term, the full gauge invariant Lagrangian
has the following accidental global charges which are conserved:
	\begin{eqnarray}
\begin{tabular}{r|ccc}
Multiplet & $\Psi^k$ & $S^{kl}$ & $\Phi^{klm}$ \\
\hline
$Q_1$ & 1 & 2 & 0 \\
$Q_2$ & 0 & 0 & 1 \\
\end{tabular}
\label{153.global}
	\end{eqnarray}
Notice that the global phase of the multiplet $\Phi$ is indeed a
global symmetry of the Lagrangian \cite{DPY91}. In addition, there is
another one, which has been labeled as $Q_1$.

Consider now a generic effective operator of the form
	\begin{eqnarray}
(\Psi)^{2f} S^{n_S} \Phi^{n_\Phi} \,,
\label{153.generic}
	\end{eqnarray}
which stands for $2f$ number of fermionic fields with upper indices,
$n_S$ number of the multiplet $S$ with upper indices etc. Each
of these numbers can be positive, negative (if the relevant multiplet
contributes a net number of lower
indices) or zero.   Thus, for example, the operator
$\Phi^{ijk}\Phi_{ijk}$ will have $n_\Phi=0$ since $\Phi^{ijk}$
contributes $n_\Phi=1$ but $\Phi_{ijk}$ contributes $n_\Phi=-1$. On
the other hand, $\Phi^{ijk}\Phi^{lmn}S_{il}S_{jm}S_{kn}$ has
$n_\Phi=2$, $n_S=-3$.

Conservation of the charges $Q_1$ and $Q_2$ tells us that, in
Eq.~(\ref{153.generic}),
	\begin{eqnarray}
f + n_S  &=& 0 \,,
\label{153.cond2} \\
n_\Phi &=& 0 \,.
\label{153.cond1}
	\end{eqnarray}
Any effective operator generated by the theory must then obey these
two conditions, and we will discuss some such operators
below.  Note that both these conditions remain unaffected if
the operator in Eq.\ (\ref{153.generic}) contains the adjoint or the
multiplet $H^{[ij]}_{[kl]}$, since they contribute equally to the
number of upper and lower indices, and since they are neutral under
the global symmetries of Eq.~(\ref{153.global}).

One general characteristic of baryon number violation in
this model can be immediately noted. As we said earlier, in order to
obtain purely fermionic operators, we need to replace the scalar
fields in Eq.\ (\ref{153.generic}) by their VEVs. From
Fig.~\ref{f:153ssb}, we note that $\Phi$ has three types of VEVs. One
of these gives $\delta B=-1$, and each of the other two give $\delta
L=1$. Let us say that in the
operator with fermionic fields only, there are $n_{\Phi_B}$ VEVs of
the first type, and $n_{\Phi_L}$ VEVs of the second kind. Since the
purely fermionic operator comes from the gauge invariant operator of
Eq.\ (\ref{153.generic}), they must obey Eq.\ (\ref{153.cond1}), which
implies $n_{\Phi_B}+n_{\Phi_L}=0$. The total violation of $B$ and $L$
in the purely fermionic operator can now be written as
	\begin{eqnarray}
\delta B = - n_{\Phi_B}, \quad
\delta L = n_{\Phi_L} = -n_{\Phi_B} \,,
	\end{eqnarray}
so that
	\begin{eqnarray}
\delta (B-L) = 0 \,.
\label{153.selrule}
	\end{eqnarray}
This immediately tells us that {\em there is no $n$-$\widehat n$
oscillations in the model.}

One comment needs to be made here.
Lepton number is not part of the gauge symmetry in SU(15). However, it
is well-defined for all components of $\Psi$. This can
be used to assign lepton number to the gauge bosons and Higgs bosons. In
other words, for any multiplet $\phi^{ij\ldots}$, we can count a lepton
number $+1$ for each occurence of the indices 13 or 14, and $-1$ for each
occurence for the index 15. Since lower indices are  complex conjugates,
they will have just the opposite assignments. It is then
easy to see that lepton number  conservation is assured in the Lagrangian
by gauge invariance,  and therefore must be violated spontaneously.
This is one characteristics of this particular version of SU(15)
models which is not shared if we break baryon number by higher rank
 multiplets, as we will see later.

\subsubsection{Proton decay operators}\label{s:153pdk}
Since proton decay requires both baryon number and lepton number
violation, and since both these violations come through VEVs of
different components of $\Phi$, we must need at least two factors of
$\Phi$ in the gauge invariant operator. Of course, one of them must
come with upper indices and the other with lower indices in order that
Eq.\ (\ref{153.cond1}) is satisfied.

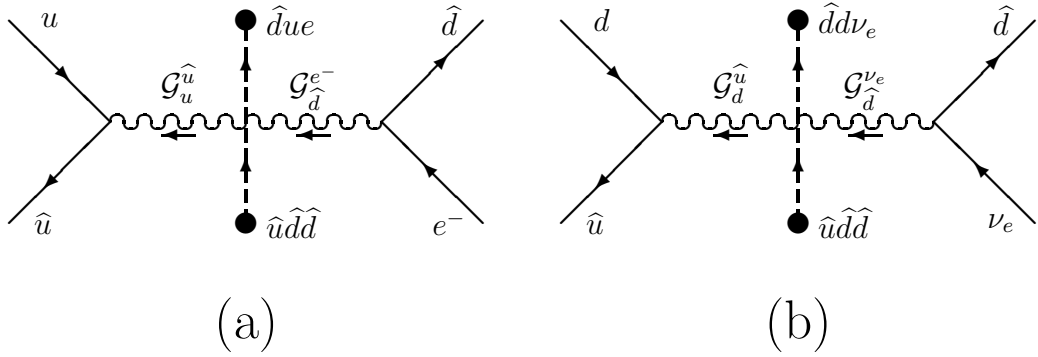
\begin{figure}
\large
\begin{center}
\begin{picture}(80,60)(-5,15)
\thicklines
\put(35,20){\makebox(0,0){{\huge (a)}}}
\put(0,65){\line(1,-1){15}}     \put(7,58){\vector(1,-1){2}}
\put(6,65){\makebox(0,0){$u$}}			
\put(0,35){\line(1,1){15}}      \put(7,42){\vector(-1,-1){2}}
\put(5,35){\makebox(0,0){$\widehat u$}}		
\put(70,65){\line(-1,-1){15}}   \put(63,58){\vector(1,1){2}}
\put(65,65){\makebox(0,0){$\widehat d$}}	
\put(70,35){\line(-1,1){15}}    \put(63,42){\vector(-1,1){2}}
\put(65,35){\makebox(0,0){$e^-$}}	
\multiput(16,50)(4,0){10}{\oval(2,2)[t]}
\multiput(18,50)(4,0){10}{\oval(2,2)[b]}
\put(25,48){\makebox(0,0){\vector(-1,0){5}}}
\put(25,54){\makebox(0,0)[b]{${\cal G}^{\widehat u}_u$}}
\put(45,48){\makebox(0,0){\vector(-1,0){5}}}
\put(45,54){\makebox(0,0)[b]{${\cal G}^{e^-}_{\widehat d}$}}
\multiput(35,35)(0,3){10}{\line(0,1){2}}
\multiput(35,43)(0,15){2}{\vector(0,1){2}}
\multiput(35,35)(0,30){2}{\circle*{3}}		
\put(38,35){\makebox(0,0)[l]{$\widehat u\widehat d\widehat d$}}
\put(38,65){\makebox(0,0)[l]{$\widehat due$}}
	\end{picture}
\begin{picture}(80,60)(-5,15)
\thicklines
\put(35,20){\makebox(0,0){{\huge (b)}}}
\put(0,65){\line(1,-1){15}}     \put(7,58){\vector(1,-1){2}}
\put(6,65){\makebox(0,0){$d$}}		
\put(0,35){\line(1,1){15}}      \put(7,42){\vector(-1,-1){2}}
\put(5,35){\makebox(0,0){$\widehat u$}}	
\put(70,65){\line(-1,-1){15}}   \put(63,58){\vector(1,1){2}}
\put(65,65){\makebox(0,0){$\widehat d$}}	
\put(70,35){\line(-1,1){15}}    \put(63,42){\vector(-1,1){2}}
\put(65,35){\makebox(0,0){$\nu_e$}}	
\multiput(16,50)(4,0){10}{\oval(2,2)[t]}
\multiput(18,50)(4,0){10}{\oval(2,2)[b]}
\put(25,48){\makebox(0,0){\vector(-1,0){5}}}
\put(25,54){\makebox(0,0)[b]{${\cal G}^{\widehat u}_d$}}		
\put(45,48){\makebox(0,0){\vector(-1,0){5}}}
\put(45,54){\makebox(0,0)[b]{${\cal G}^{\nu_e}_{\widehat d}$}}	
\multiput(35,35)(0,3){10}{\line(0,1){2}}
\multiput(35,43)(0,15){2}{\vector(0,1){2}}
\multiput(35,35)(0,30){2}{\circle*{3}}		
\put(38,35){\makebox(0,0)[l]{$\widehat u\widehat d\widehat d$}}
\put(38,65){\makebox(0,0)[l]{$\widehat dd\nu_e$}}
\end{picture}
\end{center}
\caption[]{Tree level diagram giving rise to the operator ${\cal O}_1$
of Eq.\  (\ref{153.O1}). All the indices should be considered as upper
ones, except the ones for gauge bosons $\cal G$ where upper and
lower indices have been shown explicitly.}
\label{f:153O1}
\end{figure}

\paragraph{Gauge boson mediated~:}
For gauge boson mediated proton decay for which $f=0$ in the generic
operator of Eq.\ (\ref{153.generic}),  we obtain $n_S=0$ from
Eq.\ (\ref{153.cond2}).  Indeed, there is one operator which is
consistent with all these numbers:
          \begin{eqnarray}
{\cal O}_1 =
\left[ (\overline{\Psi_L})_i \gamma^\lambda (\Psi_L)^j \right]
\left[ (\overline{\Psi_L})_k \gamma_\lambda (\Psi_L)^l \right] \;
\Phi^{ikr} \Phi_{jlr} \,,
\label{153.O1}
          \end{eqnarray}
In Fig.~\ref{f:153O1}, we have shown a tree-level diagram which gives
rise to a particular component of this operator. This is the component
with $\{ikr\} \equiv \{\widehat u \widehat d \widehat d\}$. Since the
other VEV
must have one index contracted with this one, and it has to be lepton
number violating, we have to use either $\{jlr\} =
\{\widehat due\}$ or  $\{\widehat d d\nu_e\}$. These
two possibilities are shown in Fig.\ \ref{f:153O1}a and \ref{f:153O1}b
respectively.  Once the VEVs are put in, 4-fermion
operators result, whose co-efficient can be easily computed by looking
at these diagrams:
          \begin{eqnarray}
{\cal K}_1 \simeq {g^2 M_B M_S \over M_{12}^2 M_G^2} \,.
          \end{eqnarray}
Here, the factor $g^2$ is just the gauge coupling constant coming from
two vertices with fermions. $M_B/g$ and $M_S/g$ give the VEVs, and the
four-boson vertex gives a factor $g^2$.  The denomintor comes from the
propagators of the gauge bosons.  In both diagrams, the
gauge boson coming out of the left vertex has the quantum numbers of a
diquark.  Such gauge bosons belong to the coset space
${\rm SU(12)}_q/{\rm [SU(6)}_{qL}\times {\rm SU(6)}_{qR}]$,
and acquire masses of order
of SU(12)$_q$ breaking scale, $M_{12}$. The other gauge boson which
couples a quark to a lepton belongs to the coset space
${\rm SU(15)/SU(12)}_q\times {\rm SU(3)_\ell}$, and therefore has mass
at the unification scale $M_G$.

Proton lifetime bounds now imply, from Eq.\ (\ref{Kbound}), the
constraints
          \begin{eqnarray}
M_G^2 \cdot {M_{12}^2 \over g^2 M_B M_S} > 10^{32}  \, {\rm GeV^2}\,.
\label{153.bound}
	\end{eqnarray}
Notice that $M_B$ and $M_S$ are by definition smaller than $M_{12}$,
and $g<1$.  Thus this condition can be satisfied with a low grand
unification scale.

 We now discuss the proton decay modes obtained from the
diagrams of Fig.~\ref{f:153O1}.  In the figure, we have suppressed all
generation indices of the fermions.  We now notice that there is a
property of the operator of Eq.\ (\ref{153.O1}) which forbids all
fermions to be of the same generation.  This is because
	\begin{eqnarray}
\left[ (\overline{\Psi_L})_i \gamma^\lambda (\Psi_L)^j \right]
\left[ (\overline{\Psi_L})_k \gamma_\lambda (\Psi_L)^l \right]
&=& -\left[ (\overline{\Psi_L})_i \gamma^\lambda (\Psi_L)^j \right]
\left[ (\overline{\widehat\Psi_R})^l \gamma_\lambda
(\widehat\Psi_R)_k \right] \nonumber\\
&=& 2 \left[ (\overline{\Psi_L})_i (\widehat\Psi_R)_k \right]
\left[ (\overline{\widehat\Psi_R})^l (\Psi_L)^j \right] \nonumber\\
&=& 2 \left[ (\widehat\Psi_R)_i C (\widehat\Psi_R)_k \right]
\left[  (\Psi_L)^l C  (\Psi_L)^j \right] \,. \label{153.fierz}
	\end{eqnarray}
Here, the first step is obtained by the definition of $\widehat\Psi$,
and the next one is obtained by Fierz transformation.  Since the
matrix $C$ is antisymmetric, the last form shows that the spinor
indices of $\Psi^i$ and $\Psi^k$ are antisymmetric in Eq.\
(\ref{153.O1}). Because the gauge indices are contracted with
$\Phi^{ikr}$ which is antisymmetric, the gauge indices $i$ and $k$ are
also antisymmetric.  Therefore, in order to satisfy the Fermi
principle, they must be antisymmetric in their (unshown) generation
indices. The same comment can be made about $\Psi^j$ and $\Psi^l$.
Thus, disregarding charm quark fields which will be kinematically
forbidden in proton decay, we obtain that the 4-fermion operator
generated by Fig.~\ref{f:153O1}a is
          \begin{eqnarray}
{\cal K}_1
\left[ \overline{\widehat u_L} \gamma_\lambda u_L \right] \;
\left[ \overline{\widehat s_L} \gamma^\lambda \mu_L \right] \,.
\label{153.F1a}
          \end{eqnarray}
Similarly, Fig.~\ref{f:153O1}b generates
          \begin{eqnarray}
{\cal K}_1
\left[ \overline{\widehat u_L} \gamma_\lambda d_L \right] \;
\left[ \overline{\widehat s_L} \gamma^\lambda \nu_{\mu L} \right] \,.
\label{153.F1b}
          \end{eqnarray}
The first one predicts the decay mode
	\begin{eqnarray}
p \to  \mu^+ K^0  \,,
	\end{eqnarray}
whereas the second one gives
	\begin{eqnarray}
p \to  \widehat \nu_\mu K^+  \,.
	\end{eqnarray}
Notice that these are unusual decay modes which are suppressed in
unification models like SU(5) or SO(10), although they occur in
their supersymmetric versions.  In the present case, it is predicted
in absence of supersymmetry because of the Fermi symmetry between all
particles in a generation.

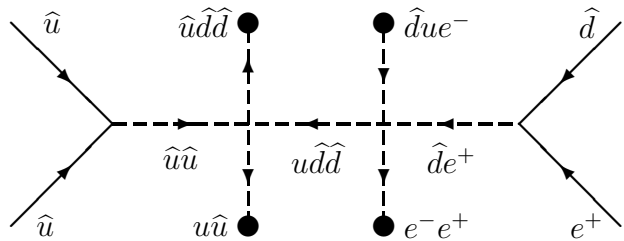
\begin{figure}
\large
\begin{center}
\begin{picture}(110,40)(-10,30)
\thicklines
\put(0,65){\line(1,-1){15}}     \put(7,58){\vector(1,-1){2}}
\put(6,65){\makebox(0,0){$\widehat u$}}		
\put(0,35){\line(1,1){15}}      \put(7,42){\vector(1,1){2}}
\put(5,35){\makebox(0,0){$\widehat u$}}		
\put(90,65){\line(-1,-1){15}}   \put(83,58){\vector(-1,-1){2}}
\put(85,65){\makebox(0,0){$\widehat d$}}		
\put(90,35){\line(-1,1){15}}    \put(83,42){\vector(-1,1){2}}
\put(85,35){\makebox(0,0){$e^+$}}		
\multiput(15,50)(3,0){20}{\line(1,0){2}}
\put(25,50){\vector(1,0){2}}
\put(25,45){\makebox(0,0){$\widehat u\widehat u$}}
\put(45,50){\vector(-1,0){2}}
\put(45,45){\makebox(0,0){$u\widehat d\widehat d$}}
\put(65,50){\vector(-1,0){2}}
\put(65,45){\makebox(0,0){$\widehat de^+$}}
\multiput(35,35)(0,3){10}{\line(0,1){2}}
\put(35,43){\vector(0,-1){2}}
\put(35,58){\vector(0,1){2}}
\multiput(55,35)(0,3){10}{\line(0,1){2}}
\put(55,43){\vector(0,-1){2}}
\put(55,58){\vector(0,-1){2}}
\multiput(35,35)(0,30){2}{\circle*{3}}		
\put(32,35){\makebox(0,0)[r]{$u\widehat u$}}
\put(32,65){\makebox(0,0)[r]{$\widehat u\widehat d\widehat d$}}
\multiput(55,35)(0,30){2}{\circle*{3}}		
\put(58,35){\makebox(0,0)[l]{$e^-e^+$}}
\put(58,65){\makebox(0,0)[l]{$\widehat due^-$}}
\end{picture}
\end{center}
\caption[]{Proton decay mediated by Higgs bosons, giving rise to the
operator in Eq.\ (\ref{153.O2}).  The notation about indices has been
explained in Fig.~\ref{f:153O1}.  One can similarly contemplate a
diagram where the VEV of the component $\Phi^{\widehat dd\nu_e}$
appears instead of $\Phi^{\widehat due}$.}\label{f:153h}
			\end{figure}

\paragraph{Higgs boson mediated~:}
In this case, we should put $f=2$ in Eq.\ (\ref{153.generic}), as
discussed in Sec.~\ref{s:general}. The constraint of Eq.\
(\ref{153.cond2}) now implies  $n_S=-2$. Operators
of this type were discussed earlier by one of us \cite{Pal92}.
 Here is one example:
     \begin{eqnarray}
{\cal O}_2 =  [( \Psi_L)^i C (\Psi_L)^j ] \, [( \Psi_L)^k C (\Psi_L)^l ]
\Phi_{ikr} \Phi^{pqr} S_{lp} S_{jq} \,.
\label{153.O2}
     \end{eqnarray}
Since the effective operator now involves two occurrences of the field
$S$ whose VEVs are of order $M_W$, any contribution coming from this
operator must have a suppression factor of $(M_W/M_G)^2$. For example,
the tree diagram of Fig.~\ref{f:153h} gives a contribution
                    \begin{eqnarray}
{\cal K}_2 \sim \left( {m_f \over M_W} \right)^2 \,
{\lambda_{S\Phi}^2 M_S M_B M_W^2 \over M_G^6} \,.
\label{153.K2}
                    \end{eqnarray}
Here, the quantity $m_f$ is the  mass of a typical fermion, and comes
from the Yukawa couplings. The quartic scalar couplings are denoted by
$\lambda_{SS}$ and  $\lambda_{S\Phi}$ in an obvious notation, and we
have assumed that all the virtual colored scalars in this diagram have
masses of order $M_G$, the largest scale in the model.
Then notice that
                    \begin{eqnarray}
{{\cal K}_2 \over {\cal K}_1} =
\lambda_{SS} \lambda_{S\Phi} \left( {m_f \over M_G} \right)^2
\left( {M_{12} \over M_G} \right)^2 \,.
\label{153.higgs/gauge}
                    \end{eqnarray}
Even if  $\lambda_{SS},\lambda_{S\Phi} \sim 1$ which is the limit
allowed by perturbative procedure, we obtain ${\cal K}_2
\ll {\cal K}_1$ since $M_{12}\leq M_G$ by definition and $m_f \ll
M_G$.

\subsection{Baryon number violated by an antisymmetric rank-4
multiplet}\label{s:154}
\subsubsection{Symmetries of the model}
                          The rank-4 antisymmetric
representation $\Delta$ was introduced by Brahmachari, Sarkar, Mann
and Steele (BSMS) \cite{BSMS92}. A possible chain of symmetry breaking
has been shown in Fig.~\ref{f:154ssb}. Notice that the symmetry
breaking above the scale $M_B$ is performed by the same VEVs as in
Fig.~\ref{f:153ssb}. At the scale $M_B$, the still unbroken gauge
group $3_{qL}2_{qL}3_{uR}3_{dR}1_B 1_{qR\Lambda} 2_{\ell L} 1_{\ell
Y}$ is broken by the VEV of the rank-4 multiplet, $\Delta^{\widehat
u\widehat u\widehat de^+}$. The $3_{uR}$ and the $3_{dR}$
subgroups combine to give the diagonal subgroup which we call
$3_{qR}$. Also, out of the three U(1) factors, one combination breaks,
leaving two unbroken ones, one of which is the hypercharge of the
standard model, and the other is called U(1)$_F$, whose generator is
proportional to $\mbox{diag}\;(9_{(6)}, -22_{(3)}, 4_{(3)},
-7_{(2)}, 14)$.

Notice that the VEV $\Delta^{\widehat u\widehat u\widehat de^+}$
breaks both baryon
number and lepton number by $-1$. Since this is the only baryon
number violating VEV in this scheme, it must appear in the gauge
invariant operator giving rise to proton decay.  Moreover, it must
appear an odd number of times. Thus, it is obvious that if the model
contains a discrete symmetry
	\begin{eqnarray}
\Delta \to - \Delta
\label{154.D-}
	\end{eqnarray}
with all other fields invariant, one cannot generate any term that
violates baryon number by an odd integer. Proton will then be
absolutely stable. This comment applies irrespective of whether proton
decay is mediated by gauge boson or Higgs boson exchange.

Even if such a symmetry is not imposed on the Lagrangian, analysis of
the full gauge invariant Lagrangian reveals accidental global
symmetries as discussed in Sec.~\ref{s:153symm},
restricting the type of potentially baryon number violating operators.
To see this, we use the notation of Eq.\ (\ref{L0L'}) and note that here
	\begin{eqnarray}
{\cal L}' = {\cal Y}
(\Psi_L)^k (\Psi_L)^l S_{kl} +
\lambda \Delta_{klmn} S_{pr} \Phi^{klp} \Phi^{mnr} +
\lambda' [\Delta \Delta \Delta \Phi]_\epsilon + \mbox{h.c.}\,,
\label{154.L'}
	\end{eqnarray}
where in the last term, the indices (not shown) are all upper indices
which are contracted by an antisymmetric $\epsilon$-tensor having 15
indices, which is indicated by the square bracket with a subscript
$\epsilon$. The first term is the Yukawa coupling term.

It is easy to see that, even with the presence of the above terms,
there is a global U(1) symmetry of the Lagrangian under which
the quantum  numbers of various multiplets are as follows:
	\begin{eqnarray}
\begin{tabular}{r|cccc}
Multiplet & $\Psi^i$ & $S^{ij}$ & $\Phi^{ijk}$ & $\Delta^{ijkl}$ \\
\hline
Charge & 1 & 2 & ${6\over 7}$ & $-{2\over 7}$\\
\end{tabular}
\label{154.global}
	\end{eqnarray}
Consider now a generic effective operator of the form
	\begin{eqnarray}
(\Psi)^{2f} S^{n_S} \Phi^{n_\Phi} \Delta^{n_\Delta} \,,
\label{154.generic}
	\end{eqnarray}
in the notation used in Eq.\ (\ref{153.generic}).
The global symmetry of Eq.\ (\ref{154.global}) implies
	\begin{eqnarray}
2f + 2n_S + {6\over 7} n_\Phi -{2\over 7} n_\Delta = 0\,.
\label{154.cond2}
	\end{eqnarray}
On the other hand, all the indices
should be contracted, which means that the total number of upper indices
should either be zero or be divisible by 15 (so that they can be
contracted by $\epsilon$-symbols). For the model of Sec.~\ref{s:153},
this condition is already contained in Eqs.\  (\ref{153.cond2}) and
(\ref{153.cond1}). Here, it produces an independent
condition\footnote{This condition is necessary, but not sufficient,
since it does not take into account the fact that the indices
to be contracted by the $\epsilon$-symbols have to be antisymmetric.}
	\begin{eqnarray}
2f + 2n_S + 3 n_\Phi + 4 n_\Delta = 15N\,,
\label{154.cond1}
	\end{eqnarray}
where $N$ is an integer, denoting the number of times a vertex
involving the $\epsilon$-symbol appears in the diagram giving rise to
the operator of Eq.\ (\ref{154.generic}). As noted in
Sec.~\ref{s:153symm}, both conditions remain unaffected if
the operator in Eq.\ (\ref{154.generic}) contains the adjoint or the
antisymmetric bi-adjoint.

The solution of Eqs.\  (\ref{154.cond2}) and (\ref{154.cond1})
can be written as:
	\begin{eqnarray}
f+n_S= n_\Delta -3N, \quad n_\Phi=7N - 2n_\Delta \,.
\label{154.solution}
	\end{eqnarray}
Let us now check what the above solution means for the violation of
baryon and lepton numbers. Baryon number, as noted before, is part of
the gauge symmetry and is broken only spontaneously through the VEV of
$\Delta$. Thus, clearly,
	\begin{eqnarray}
\delta B = - n_\Delta \,.
\label{154.deltaB}
	\end{eqnarray}
On the other hand, lepton number violation comes from three different
sources:
	\begin{itemize}
\item each VEV of $\Delta$ (with upper indices) gives $\delta L=-1$;
\item each VEV of $\Phi$ (with upper indices) induces $\delta L=1$;
\item each occurence of a term with an $\epsilon$-symbol will have 15
upper indices which are all different, contributing to an
explicit violation $\delta L=1$ in the unbroken Lagrangian.
	\end{itemize}
Taking all these contributions, we can write
	\begin{eqnarray}
\delta L = -n_\Delta + n_\Phi + N \,.
\label{154.deltaL}
	\end{eqnarray}
Using Eqs.\  (\ref{154.solution}-\ref{154.deltaL}), we therefore
finally obtain
	\begin{eqnarray}
\delta (3B-L) = -8N \,,
\label{154.selrule}
	\end{eqnarray}
which is the selection rule for this model.
Immediately, it tells us that in this model,
{\em there cannot be any neutron-antineutron oscillations}.

\subsubsection{Proton decay operators}
                           Specializing to the simplest case when
$N=0$, Eq.\ (\ref{154.selrule}) tells us that $3B-L$
is conserved, which means that there will be three leptons in the
final state for proton decay. This cannot occur with four fermionic
fields only, since three of these fields must be quark/antiquark
fields in order to obtain a $\delta B=-1$ operator. For other values
of $N$, one needs even higher number of leptons/antileptons in the
final state, which cannot be accommodated in a 4-fermion operator for
the same reason. Thus, we conclude
that in this model, there is no proton decay operator with four
fermionic field operators. The result is true for operators mediated
by gauge or Higgs bosons.

The lowest dimensional operators will thus have six fermionic fields.
They can have $f=3$ where all the indices are upper.  Alternatively,
they may have $f=1$ where two of the fields have lower indices, but
the other four have upper ones. Of course, one can similarly have
$f=-3$ and $f=-1$.

Among these possibilities, $f=1$ can yield a solution to Eq.\
(\ref{154.solution}) with smallest number of scalar fields, given by
	\begin{equation}
f=1, \quad n_S =0, \quad n_\Phi=-2, \quad n_\Delta=1.
\label{154.simplest}
	\end{equation}
An operator of this type is:
	\begin{eqnarray}
{\cal O} =
\left[ (\overline{\Psi_L})_i \gamma_\lambda (\Psi_L)^j \right] \,
\left[ (\overline{\Psi_L})_k \gamma^\lambda (\Psi_L)^l \right] \,
\left[ (\Psi_L)^p C (\Psi_L)^r \right] \;
\Delta^{ikab} \Phi_{ajr} \Phi_{plb}  \,.
\label{154.O1}
	\end{eqnarray}
We show in
Fig.~\ref{f:154pdk} how this operator can arise at the
tree level.  The amplitude of the purely fermionic operators can be easily
determined.  Assuming the scalar
interaction couplings to be of order unity, we obtain
	\begin{eqnarray}
{\cal K}_{(6)} \sim {g M_G M_B M_S \over M_{12}^2 M_G^2
M_{\widehat due^-e^+}^2 M_{e\nu}^2} \,,
\label{154.bound}
	\end{eqnarray}
where the last two factors in the denominator represent the masses of
the internal Higgs boson lines. Of these, the former one is a colored
boson, whose mass is expected to be of order $M_G$. But the latter one
is uncolored, whose mass we keep as an unknown. Experimental bounds,
however, tell us that, being a charged scalar, its mass cannot be
much less than 100 GeV. Thus, if this operator contributes to proton
decay, using Eq.\ (\ref{K2nbound}), we can rewrite Eq.\
(\ref{154.bound}) as
	\begin{eqnarray}
M_G^3 > 10^{28} \, {\rm GeV}^3 \; \times \left( {gM_BM_S \over M_{12}^2}
\right) \cdot \left( {100\, {\rm GeV} \over M_{e\nu}}
\right)^2 \,.
	\end{eqnarray}
Since $M_{12}>M_B,M_S$ by definition and $g<1$, this bound can be
satisfied for any unification scale larger than about $10^9$
GeV.

However, there is a subtle reason why this operator cannot contribute
to proton decay.  In order to accommodate baryon number violation, the
indices on the fields $\Delta$ must be $\widehat u\widehat u\widehat
de^+$ in any permutation. Now, these indices contract either with the
indices of $\Phi$, or those of $\overline\Psi$. But $\Phi$ does not
have any VEV which contains the index $\widehat u$. Thus, both the
indices on $\overline\Psi$ have to be $\widehat u$ indices. However,
as argued in connection with Eq.\ (\ref{153.fierz}), the fields
$(\overline{\Psi_L})_i$ and $(\overline{\Psi_L})_k$ must come
from different generations.  Therefore, one of them must be the charm
quark and therefore proton decay is kinematically forbidden from this
operator.

To get out of this impasse, one can use a slightly modified operator:
	\begin{eqnarray}
{\cal O}' =
\left[ (\overline{\Psi_L})_i \gamma_\lambda (\Psi_L)^j \right] \,
\left[ (\overline{\Psi_L})_k \gamma^\lambda (\Psi_L)^l \right] \,
\left[ (\Psi_L)^p C (\Psi_L)^r \right] \;
T^i{}_m \Delta^{mkab} \Phi_{ajr} \Phi_{plb}  \,.
\label{154.O'}
	\end{eqnarray}
This is still an  operator of the type of Eq.\ (\ref{154.simplest}),
but now the gauge indices $i$ and $k$ are not antisymmetric, and
therefore $(\overline{\Psi_L})_i$ and $(\overline{\Psi_L})_k$ can
refer to fields from the same generation.  A diagram for this operator
can be obtained from Fig.~\ref{f:154pdk}
by attaching an adjoint Higgs boson to any line which
carries at least one SU(12)$_q$ index.  This will provide further
suppression to the 4-fermion operators since the extra propagator is
expected to have a mass $M_G$, but the largest VEV available for the
adjoint multiplet is at the scale $M_{12}$.  The quark level
transition induced by this operator is
$ue^-e^-\nu \to \widehat u\widehat u$, which implies a decay mode
	\begin{eqnarray}
p \to \pi^- e^+ e^+ \widehat\nu_e \,.
	\end{eqnarray}

\subsection{Baryon number violated by an antisymmetric rank-5
multiplet}\label{s:155}
          In an early  paper, Frampton and Kephart \cite{FrKe90}
discussed baryon number violation by the VEV of an antisymmetric
rank-5  multiplet $J^{[ijklm]}$. Although less economical than
the ones discussed above, we include this possibility for the sake of
completeness. Fig.~\ref{f:155ssb} gives a chain involving this rank-5
multiplet. The VEV that breaks $U(1)_B$ has the gauge
transformation properties of $\widehat{d} \widehat{d} \widehat{d}
\nu_e e^-$, i.e, it has $B=-1$, $L=2$.  Notice also that since this
VEV does not involve
both $\widehat u$ and $\widehat d$ type indices, it cannot break
$3_{uR}3_{dR}$ part of the symmetry to $3_{qR}$, as is done in the
models described earlier. Therefore, this symmetry breaking is
performed by a VEV in the antisymmetric bi-adjoint $H^{[ij]}_{[kl]}$.
 This multiplet certainly has a component which is the
antisymmetric bi-adjoint of the subgroup $6_{qR}$ and singlet under
the rest. This part is a 189-dimensional representation of SU(6) which
has a component that transforms like $(8,8,0)$ under its subgroup
$3_{uR}3_{dR}1_{qR\Lambda}$. A VEV here would perform the desired
symmetry breaking.    On the other hand, one
now does not need the adjoint to break the $6_{qR}$ subgroup, since
the baryon number violating VEV itself performs the job. In fact, the
VEV  $\langle J^{\widehat{d} \widehat{d} \widehat{d}
\nu_e e^-}\rangle$ also breaks the leptonic subgroup $3_\ell$ to
$2_{\ell L}$, and the leptonic and quark hypercharges combine to the
total hypercharge of the standard model.

In this case, using the notation introduced earlier, we obtain
	\begin{eqnarray}
{\cal L}' = {\cal Y} (\Psi_L)^k (\Psi_L)^l S_{kl} + \mu
[JJJ]_\epsilon  + \mbox{h.c.} \,.
\label{155.L'}
	\end{eqnarray}
Obviously, the entire Lagrangian respects the discrete symmetry
	\begin{eqnarray}
J \to e^{2\pi i/3} \, J
\label{155.discrete}
	\end{eqnarray}
with all other fields neutral. This is a $Z_3$ symmetry. The number of
$J$ fields in any effective operator arising in this model must then
be a multiple of 3.  Since baryon
number violation comes from the VEV of $J$ only, in purely fermionic
operators we will have
	\begin{eqnarray}
\left| \delta B \right| = 3N
	\end{eqnarray}
for some integer $N$. Therefore, {\em neither proton decay nor
neutron-antineutron oscillation is possible in this
model.}\footnote{Indeed, the baryon number violating diagram given by
Frampton and Kephart \cite{FrKe90} for this model has $\delta B=3$, as
was noted by one of us earlier \cite{Pal92}.}  Notice
that this conclusion is reached only from the accidental
symmetries present in the full gauge invariant Lagrangian.

\subsection{Introduction of antisymmetric Yukawa
couplings}\label{s:15A}
                   So far, we have assumed that the only
Higgs bosons which can couple to fermions belong to the symmetric
rank-2 multiplet  $S^{\{ij\}}$.
The situation changes if, in addition there is also the multiplet
$A^{[ij]}$ which couples
antisymmetrically. In this case, some of the symmetries described in
the above sections may be broken explicitly and hence more baryon number
violating processes may be allowed.

 For the model of Sec.~\ref{s:153}, such is not the case.
We still have the condition in Eq.\ (\ref{153.cond1}), which leads to
$B-L$ conservation. However, in the model of Sec.~\ref{s:154}, there
is an important change. This is because, with the introduction of the
multiplet $A$, there are the following new terms which are allowed in
${\cal L}'$:
	\begin{eqnarray}
{\cal L}'_A = {\cal Y}_A (\Psi_L)^k (\Psi_L)^l A_{kl} + \mu A_{ij}
A_{kl} \Delta^{ijkl} +\mbox{h.c.} \,.
\label{15A4.L'}
	\end{eqnarray}
There is now no way that one can assign a quantum number of $A$ which
keeps the symmetry of Eq.\ (\ref{154.global}).  Thus, one can have the
following operator:
	\begin{eqnarray}
{\cal O} =  [( \Psi_L)^i C (\Psi_L)^j ] \, [( \Psi_L)^k C (\Psi_L)^l ]
\Delta_{ijkl} \,.
\label{15A4.O}
	\end{eqnarray}
In Fig.~\ref{f:15A4}, we show how this can be generated through the
interactions appearing in Eq.\ (\ref{15A4.L'}). In the figure, we
suppressed the generation indices. Turning to Eq.\ (\ref{15A4.O}), we
see that since the gauge group indices $i$ and $j$ appear in
antisymmetric combination in $\Delta^{ijkl}$, and since the matrix $C$
is antisymmetric, the generation indices for the two fermionic fields
in the first bilinear must be antisymmetric in order to maintain Fermi
symmetry. The same can be said about the fermionic fields in the other
bilinear. Thus, the quark level operator coming from Fig.~\ref{f:15A4}
is $[\widehat u C \mu^+]\,[\widehat u C \widehat s]$, which gives
rise to a proton decay mode
	\begin{eqnarray}
p \to \mu^+ K^0 \,.
	\end{eqnarray}
 The amplitude for the 4-fermion operator is given by
	\begin{eqnarray}
{\cal K} \sim {\cal Y}_A^2 {\mu M_B \over M_G^4} \,,
	\end{eqnarray}
assuming, once again, that the colored Higgs bosons have masses of
order $M_G$. The quantity ${\cal Y}_A$ in this formula stands
symbolically for two factors of the Yukawa coupling with the multiplet
$A$. Since the antisymmetric Yukawa couplings, if any, are expected to
be smaller than the symmetric ones, and since $M_B<M_G$ by definition,
this again shows the suppression of proton decay rate.

For the model of Sec.~\ref{s:155}, the changes are more dramatic.
Here, the extra terms can appear in ${\cal
L}'$ due to the introduction of $A$ are given by
	\begin{eqnarray}
{\cal L}'_A = {\cal Y}_A (\Psi_L)^k (\Psi_L)^l A_{kl} +
\lambda_{JJ\Phi A} [JJ\Phi A]_\epsilon + \mu'
J^{ijklm} \Phi_{klm} A_{ij} + \mbox{h.c.}\,.
\label{15A5.L'}
	\end{eqnarray}
There still is a $Z_3$ symmetry in the full Lagrangian, defined as
follows:
	\begin{eqnarray}
\begin{tabular}{r|ccccc}
Multiplet & $\Psi^k$ & $S^{kl}$ & $A^{kl}$ & $\Phi^{klm}$
& $J^{ijklm}$ \\
\hline
$Z_3$  charge & 1 & 2 & 2 & 2 & 1 \\
\end{tabular} .
\label{15A5.Z3}
	\end{eqnarray}
Consider now a generic effective operator of the form
	\begin{eqnarray}
(\Psi)^{2f} S^{n_S} A^{n_A} \Phi^{n_\Phi} J^{n_J}
\label{15A5.generic}
	\end{eqnarray}
in the notation used before. Using the $Z_3$ symmetry and
the requirement that all indices must be contracted, we obtain
the following conditions:
	\begin{eqnarray}
2f + 2 (n_S + n_A) + 2 n_\Phi + n_J &=& 3N \,, \label{15A5.symmcond}\\
2f + 2 (n_S + n_A) + 3 n_\Phi + 5 n_J &=& 15N'\,, \label{15A5.matchcond}
	\end{eqnarray}
where $N$ and $N'$ are both integers. Thus,
	\begin{eqnarray}
n_\Phi = 15N' - 3N - 4n_J \,, \quad 2f + 2(n_S + n_A) = 9N - 30N' +
7n_J \,.
\label{15A5.solution}
	\end{eqnarray}
Following arguments similar to those in Sec.~\ref{s:154}, we now obtain
	\begin{eqnarray}
\delta B &=& -n_J \,, \label{15A5.deltaB}\\
\delta L &=& 2n_J + n_\Phi + N' \,, \label{15A5.deltaL}
	\end{eqnarray}
so that, using Eq.\ (\ref{15A5.solution}), we obtain
	\begin{eqnarray}
\delta (2B -L) = 3N - 16N' \,,
\label{15A5.selrule}
	\end{eqnarray}
which is the selection rule in this case.

For proton decay which requires $n_J=1$,
notice that the integer $N$ must be odd because of its definition in
Eq.\ (\ref{15A5.symmcond}).  The solution of Eq.\ (\ref{15A5.solution})
involving minimum number of scalar fields is now given by $N=-1$,
$N'=0$, i.e.,
	\begin{eqnarray}
n_\Phi = -1, \quad f + n_S + n_A = -1 \,.
	\end{eqnarray}
Using Eqs.\  (\ref{15A5.deltaB}) and (\ref{15A5.deltaL}), it is now easy
to see that in this case, proton decay operators will satisfy the
selection rule $\delta (B+L)=0$.

Example of a gauge boson mediated diagram of proton decay is provided
in Fig.~\ref{f:15A5g}, which has $f=0$ and $n_A=-1$. The operator
here has the form
          \begin{eqnarray}
{\cal O}_1 =
\left[ (\overline{\Psi_L})_i \gamma^\lambda (\Psi_L)^j \right] \;
\left[ (\overline{\Psi_L})_k \gamma_\lambda (\Psi_L)^l \right] \;
J^{ikmnp} \Phi_{jmn} A_{lp} \,,
\label{15A5.O1}
          \end{eqnarray}
and Fig.~\ref{f:15A5g} shows  how it can be generated at
the tree level. Because of the Fierz transformation property shown in
Eq.\ (\ref{153.fierz}), the generation indices of $\Psi^i$ and
$\Psi^k$ must be different here. But the same cannot be said about
$\Psi^j$ and $\Psi^l$ since their gauge indices are not antisymmetric.
Thus, the quark-level transition obtained from Fig.~\ref{f:15A5g} is
$de^+ \to \widehat s\widehat d$. This implies a proton decay mode
	\begin{eqnarray}
p \to \pi^+ K^+ e^- \,,
\label{15A5.mode}
	\end{eqnarray}
which conserves $B+L$, as argued before on general grounds.
The coefficient of the 4-fermion operator is given by
	\begin{eqnarray}
{\cal K}_1 \simeq {\mu' M_B M_S M_W \over M_{12}^2 M_G^4} \,,
	\end{eqnarray}
assuming that the colored scalar internal line has a mass of order
$M_G$. Once again, since $M_B$, $M_S$ and $M_W$ are each smaller than
either $M_{12}$ or $M_G$ by definition, a low unification scale is
allowed.

A Higgs boson mediated diagram, with $f=-2$ and $n_A=1$, was given in
Ref. \cite{Pal92}. The operator responsible for this is:
          \begin{eqnarray}
{\cal O}_2 =
\left[ ({\Psi_L})^i C (\Psi_L)^j \right] \;
\left[ ({\Psi_L})^k C (\Psi_L)^l \right] \;
J_{ijkqr} \Phi^{pqr} A_{lp} \,.
\label{15A5.O2}
          \end{eqnarray}
It apparently looks like it has the same VEVs as the operator in
Eq.\ (\ref{15A5.O1}). But this need not be the case, as seen from
Fig.~\ref{f:15A5h}. Here, one can use the VEV of $\Phi$ which occurs
at the scale $M_G$. Thus, we obtain for the strength of the 4-fermion
operator
	\begin{eqnarray}
{\cal K}_2 \sim \left( {m_f \over M_W} \right)^2 {\mu' M_B M_W \over
M_G^5} \,.
	\end{eqnarray}
Depending on the magnitude of the scales $M_S$ and $M_{12}$,
this may or may not dominate over the gauge boson mediated decay.
 The
decay mode is the same as that given in Eq.\ (\ref{15A5.mode}) since
$\Psi^i$ and $\Psi^j$ have to belong to different generations in the
operator of Eq.\ (\ref{15A5.O2}).

It must also be noticed that, unlike the previous models,
neutron-antineutron oscillations are not ruled out in this model.
However, it is very suppressed. This can be seen from Eq.\
(\ref{15A5.selrule}), where we can put $\delta B=2$ and $\delta L=0$
as is necessary for neutron-antineutron oscillations. The simplest
solution for this situation is obtained when $N=-4$, $N'=-1$, which
means
	\begin{eqnarray}
n_J = -2, \quad n_\Phi= 5, \quad f+n_S+n_A = -10 \,.
	\end{eqnarray}
Obviously, it is a very high dimensional operator, so we will ignore
it.

\section{Scenarios of baryon number violation in SU(16)
models}\label{s:16}
                  It might seem that SU(16) scenarios of baryon number
violation should look similar to the SU(15) ones, since the groups are
not all that different.  There are, however, some important
differences, which should carefully be taken into account. The first
is that baryon number violation occurs spontaneously and therefore is
sensitive to the choice of the Higgs sector, as amply demonstrated in
Sec.~\ref{s:15}. Being a larger group, SU(16) in general requires more
VEVs to break it down to the standard model gauge group, which affect
the operator analysis. Secondly, baryon number processes like proton
decay involves lepton number violation as well, and the latter is very
different in the groups SU(15) and SU(16). The reason is that in
SU(16) lepton number is part of the gauge symmetry and can be violated
only spontaneously. This is a difference from the SU(15) models where
the Lagrangian can violate lepton number. Thirdly, the symmetric
rank-2 multiplet of SU(16), unlike its SU(15) version, can have a
lepton number violating VEV $S^{\widehat\nu\widehat\nu}$, which does
not
violate the symmetries of the standard model. This VEV can give
neutrinos a Majorana mass at the tree level.  To keep our discussion
simple, we will neglect this VEV.

\subsection{Baryon number violated by an antisymmetric rank-4
multiplet}\label{s:164}
\subsubsection{Symmetries of the model}\label{s:164symm}
                   Breaking SU(15) down to
SU(12)$_q\times$SU(3)$_\ell$ requires the VEV of an antisymmetric
rank-3 multiplet. Similarly, breaking SU(16) down to
SU(12)$_q\times$SU(4)$_\ell$ requires the VEV of an antisymmetric
rank-4 multiplet $\Delta$.  Therefore, it is reasonable to try
to see if there are suitable VEVs in the multiplet $\Delta$, the
adjoint $T^i{}_j$ and
the bi-adjoint $H^{[ij]}_{[kl]}$ which can break the grand unification
symmetry down to the symmetry of the standard model. In
Fig.~\ref{f:164ssb}, we show how it can be done. At the weak scale,
the symmetry is broken by the rank-2 symmetric multiplet $S^{\{ij\}}$,
as before.

For this model, we find
	\begin{eqnarray}
{\cal L}' = {\cal Y}  (\Psi_L)^i (\Psi_L)^j S_{ij} +
\lambda' [\Delta \Delta \Delta \Delta]_\epsilon + \mbox{h.c.}\,.
	\end{eqnarray}
It is easy to see that it has an accidental global U(1)$\times Z_4$
symmetry, under which the charges of various multiplets are as
follows:
	\begin{eqnarray}
\begin{tabular}{r|ccc}
Multiplet & $\Psi^i$ & $S^{ij}$ & $\Delta^{ijkl}$ \\
\hline
U(1) charge & 1 & 2 & 0 \\
$Z_4$ charge & 0 & 0 & 1 \\
\end{tabular}
\label{164.u1z4}
	\end{eqnarray}
Considering now a generic effective operator of the form
	\begin{eqnarray}
(\Psi)^{2f} S^{n_S} \Delta^{n_\Delta} \,,
\label{164.generic}
	\end{eqnarray}
The requirements of the U(1)$\times Z_4$ symmetry and of the
contraction of all indices give the following conditions:
	\begin{eqnarray}
2f + 2 n_S  &=& 0 \,, \label{164.u1cond}\\
n_\Delta &=& 4N \,, \label{164.z4cond}
	\end{eqnarray}
for some integer $N$. Notice that both baryon number and lepton number
violation come from only one VEV, viz., $\langle\Delta^{\widehat
u\widehat d\widehat d\widehat\nu}\rangle$. This VEV gives
$\delta B=-1$, $\delta L=-1$. Thus, in this model,
	\begin{eqnarray}
\delta (B-L) = 0 \,.
\label{164.selrule}
	\end{eqnarray}
Once again, {\em neutron-antineutron oscillation is not possible in
this model.} The possibilities of proton decay are discussed below.

\subsubsection{Proton decay operators}
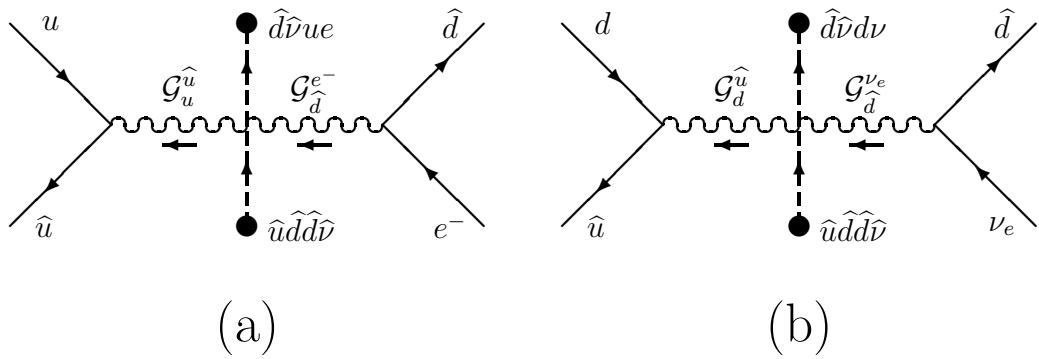
\begin{figure}
\large
\begin{center}
\begin{picture}(80,60)(-5,15)
\thicklines
\put(35,20){\makebox(0,0){{\huge (a)}}}
\put(0,65){\line(1,-1){15}}     \put(7,58){\vector(1,-1){2}}
\put(6,65){\makebox(0,0){$u$}}		
\put(0,35){\line(1,1){15}}      \put(7,42){\vector(-1,-1){2}}
\put(5,35){\makebox(0,0){$\widehat u$}}	
\put(70,65){\line(-1,-1){15}}   \put(63,58){\vector(1,1){2}}
\put(65,65){\makebox(0,0){$\widehat d$}}	
\put(70,35){\line(-1,1){15}}    \put(63,42){\vector(-1,1){2}}
\put(65,35){\makebox(0,0){$e^-$}}	
\multiput(16,50)(4,0){10}{\oval(2,2)[t]}
\multiput(18,50)(4,0){10}{\oval(2,2)[b]}
\put(25,47){\makebox(0,0){\vector(-1,0){5}}}
\put(25,54){\makebox(0,0)[b]{${\cal G}^{\widehat u}_u$}}
\put(45,47){\makebox(0,0){\vector(-1,0){5}}}
\put(45,54){\makebox(0,0)[b]{${\cal G}^{e^-}_{\widehat d}$}}
\multiput(35,35)(0,3){10}{\line(0,1){2}}
\multiput(35,43)(0,15){2}{\vector(0,1){2}}
\multiput(35,35)(0,30){2}{\circle*{3}}		
\put(43,35){\makebox(0,0){$\widehat u\widehat d\widehat
d\widehat\nu$}}
\put(43,65){\makebox(0,0){$\widehat d\widehat\nu ue$}}
	\end{picture}
\begin{picture}(80,60)(-5,15)
\thicklines
\put(35,20){\makebox(0,0){{\huge (b)}}}
\put(0,65){\line(1,-1){15}}     \put(7,58){\vector(1,-1){2}}
\put(6,65){\makebox(0,0){$d$}}		
\put(0,35){\line(1,1){15}}      \put(7,42){\vector(-1,-1){2}}
\put(5,35){\makebox(0,0){$\widehat u$}}	
\put(70,65){\line(-1,-1){15}}   \put(63,58){\vector(1,1){2}}
\put(65,65){\makebox(0,0){$\widehat d$}}	
\put(70,35){\line(-1,1){15}}    \put(63,42){\vector(-1,1){2}}
\put(65,35){\makebox(0,0){$\nu_e$}}	
\multiput(16,50)(4,0){10}{\oval(2,2)[t]}
\multiput(18,50)(4,0){10}{\oval(2,2)[b]}
\put(25,47){\makebox(0,0){\vector(-1,0){5}}}
\put(25,54){\makebox(0,0)[b]{${\cal G}^{\widehat u}_d$}}
\put(45,47){\makebox(0,0){\vector(-1,0){5}}}
\put(45,54){\makebox(0,0)[b]{${\cal G}^{\nu_e}_{\widehat d}$}}
\multiput(35,35)(0,3){10}{\line(0,1){2}}
\multiput(35,43)(0,15){2}{\vector(0,1){2}}
\multiput(35,35)(0,30){2}{\circle*{3}}		
\put(43,35){\makebox(0,0){$\widehat u\widehat d\widehat d\widehat\nu$}}
\put(43,65){\makebox(0,0){$\widehat d\widehat\nu d\nu$}}
	\end{picture}
\end{center}
\caption[]{Tree level diagram giving rise to the operator ${\cal O}_1$
of Eq.\  (\ref{164.O1}).}
\label{f:164O1}
\end{figure}

             Obviously, the simplest solution to Eqs.\
(\ref{164.u1cond}) and (\ref{164.z4cond}) are given by
	\begin{eqnarray}
f=n_S=n_\Delta = 0 \,.
	\end{eqnarray}
An operator of this type is:
	\begin{eqnarray}
{\cal O}_1 =
\left[ (\overline{\Psi_L})_i \gamma^\lambda (\Psi_L)^j \right] \;
\left[ (\overline{\Psi_L})_k \gamma_\lambda (\Psi_L)^l \right] \;
\Delta^{ikpq} \Delta_{pqjl} \,.
\label{164.O1}
	\end{eqnarray}
This can give gauge-boson mediated proton decay, as shown in
Fig.~\ref{f:164O1}. Notice that this diagram is very similar to
Fig.~\ref{f:153O1}. The analysis is also the same, leading to the
constraint in Eq.\ (\ref{153.bound}). In fact, one can also show
that Higgs boson mediated diagrams, having $n_S=2$, will be suppressed
in this model, as shown in Eq.\ (\ref{153.higgs/gauge}).

\subsection{Baryon number violated by an antisymmetric rank-3
multiplet}\label{s:163}
           Deshpande, Keith and Pal \cite{DKP93} advocated a model
where baryon number symmetry is violated by a VEV of an antisymmetric
rank-3 multiplet as in Sec.~\ref{s:153}. In Fig.~\ref{f:163ssb}, we
show this chain with some slight modifications which helps eliminate a
fundamental Higgs multiplet which was used by them.

With the introduction of the multiplet $\Phi$, there is one more term
in $\cal L'$:
	\begin{eqnarray}
{\cal L}' = {\cal Y}  (\Psi_L)^i (\Psi_L)^j S_{ij} +
\lambda \Delta_{klmn} S_{pr} \Phi^{klp} \Phi^{mnr} +
\lambda' [\Delta \Delta \Delta \Delta]_\epsilon + \mbox{h.c.} \,.
\label{163.L'}
	\end{eqnarray}
However, we now have an accidental U(1)$\times Z_8$
symmetry, with the
following charge assignments:
	\begin{eqnarray}
\begin{tabular}{r|cccc}
Multiplet & $\Psi^i$ & $S^{ij}$ & $\Phi^{ijk}$ & $\Delta^{ijkl}$ \\
\hline
U(1) charge  & 1 & 2 & 1 & 0 \\
$Z_8$ charge & 0 & 0 & 1 & 2 \\
\end{tabular}
\label{163.u1z4}
	\end{eqnarray}
So now, the generic effective operator of the form
	\begin{eqnarray}
(\Psi)^{2f} S^{n_S} \Phi^{n_\Phi} \Delta^{n_\Delta}
\label{163.generic}
	\end{eqnarray}
is subject to the following constraints:
	\begin{eqnarray}
2f + 2 n_S + n_\Phi &=& 0 \,, \label{163.u1cond}\\
n_\Phi + 2n_\Delta &=& 8N  \,, \label{163.z4cond}
	\end{eqnarray}
for some integer $N$. The first of these equations now implies that
$n_\Phi$ must be even.  The simplest solution to
these conditions is given by all $n$'s being zero, which gives the
operator of Eq.\ (\ref{153.O1}), and the phenomenological conclusions
are the same as in Sec.~\ref{s:153}.

\section{Conclusions}
         We have analyzed a variety of symmetry breaking chains within
the gauge groups SU(15) and SU(16) in which the grand unified gauge
group breaks to ``un-unified'' subgroups under which quarks and
leptons have separate symmetries.  As mentioned in the Introduction,
such chains are interesting because some of them are known to predict
low unification scales, sometimes as low as $10^8$ GeV.  Our analysis
shows that low unification scale is not phenomenologically ruled
out in these models because proton decay operators are very
suppressed. The amount of suppression, of course, depends on the Higgs
boson sector of the models and therefore varies from one model to
another.   We have also shown that in these models, since
operator analysis can be performed on the full gauge invariant
operators, and since such operators have a large Fermi symmetry, the
proton cannot decay into non-strange hadrons.  Such modes are
preferred in supersymmetric SU(5) or SO(10) models, but here we obtain
this conclusion without any supersymmetry in our models.  In fact,
inclusion of supersymmetry in SU(15) or SU(16) models typically make
the unification scales large \cite{BrSa93}. In that case, with all the
suppression mentioned in this paper, proton decay should be
unobservably slow.

One remarkable result that comes out from our analysis is that, most
of these models contain accidental global or discrete symmetries.  Of
course, these symmetries depend on the Higgs boson contents of the
model, much like the $B-L$ conservation in the simplest SU(5) unification
model.   Such symmetries provide selection rules to  baryon number
violating processes.  For example, neutron-antineutron oscillation is
strictly forbidden in most of the models, as we pointed out.  Of
course, one can always further complicate the models, using more Higgs
boson multiplets than are necessary for breaking the symmetries.
Presence of these multiplets will explicitly break some or all of the
accidental symmetries that we discovered, and therefore will allow
more baryon number violating processes.   We provided examples of this
by introducing, in Sec.~\ref{s:15A}, the antisymmetric rank-2 tensor
which couples to fermions.  Another example could be a multiplet
$X^{\{[ijk][lmn]\}}$, which exists in the symmetric part of the tensor
product of two antisymmtric rank-3 multiplets.  Once this multiplet is
introduced, one can show that neutron-antineutron oscillations become
allowed in most cases through the operator
	\begin{eqnarray}
[( \Psi_L)^i C (\Psi_L)^j ] \, [( \Psi_L)^k C (\Psi_L)^l ]
\, [( \Psi_L)^m C (\Psi_L)^n ] X_{\{[ijk][lmn]\}} \,.
	\end{eqnarray}
Our analysis, however, deals mostly with ``minimal'' models
in the sense that
we do not introduce any Higgs boson multiplets which are not necessary
for symmetry breaking.

Our results can also be used to look for other baryon and lepton
number violating processes in these models. For example, using Eqs.\
(\ref{153.selrule}), (\ref{154.selrule}) and (\ref{164.selrule}), we
can conclude that
neutrinos cannot have any Majorana mass in the SU(15) models of Secs.\
\ref{s:153} and \ref{s:154}, as well as in the SU(16) model of
Sec.~\ref{s:164}.  For the first of these models, this
result was derived by earlier authors \cite{DPY91}, but for the other
models, the result is new.  This and other new
results can be readily derived from the accidental symmetries that we
have discovered in this article for various models of interest.

\paragraph*{Acknowledgements~:} The work of PBP was supported by the
Department of Energy of the United States. US would like to thank
Professor Patrick J. O'Donnell
for arranging his visit to the University of Toronto and the NSERC
of Canada for an International Scientific Exchange Award. We thank
Daniel Wyler for reading the manuscript carefully and making insightful
comments.


\newpage

	\begin{figure}
\begin{center}
\Large
\begin{tabular} {rcl}
\multicolumn{3}{c}
    {\framebox{$15$}}\\
$M_G$ & $\Huge \Downarrow$ & $\langle \Phi^{\nu_ee^-e^+}\rangle$\\
\multicolumn{3}{c}
    {\framebox{$12_q 3_\ell$}}\\
$M_{12}$ & $\Huge \Downarrow$ & $\langle 1_{(6)}, -1_{(6)},
    0_{(3)}\rangle$\\
\multicolumn{3}{c}
    {\framebox{$6_{qL}6_{qR}1_B 3_\ell$}}\\
$M_{6qL}$ & $\Huge \Downarrow$ & $\langle H^{[ij]}_{[kl]}\rangle$\\
\multicolumn{3}{c}
    {\framebox{$3_{qL}2_{qL}6_{qR}1_B 3_\ell$}}\\
$M_{6qR}$ & $\Huge \Downarrow$
                 & $\langle 0_{(6)},1_{(3)}, -1_{(3)}, 0_{(3)}\rangle$\\
\multicolumn{3}{c}
    {\framebox{$3_{qL}2_{qL}3_{uR}3_{dR}1_B 1_{qR\Lambda} 3_\ell$}}\\
$M_{3\ell}$ & $\Huge \Downarrow$ & $\langle 0_{(12)},1_{(2)}, -2\rangle$\\
\multicolumn{3}{c}
    {\framebox{$3_{qL}2_{qL}3_{uR}3_{dR}1_B 1_{qR\Lambda}
                  2_{\ell L} 1_{\ell Y}$}}\\
$M_B$ & $\Huge \Downarrow$ & $\langle \Phi^{\widehat u \widehat d
\widehat d}\rangle$\\
\multicolumn{3}{c}
    {\framebox{$3_{qL}2_{qL}3_{qR} 1_{qY}
                  2_{\ell L} 1_{\ell Y}$}}\\
$M_S$ & $\Huge \Downarrow$ & $\langle \Phi^{\widehat d
(ue-d\nu_e)}\rangle$\\
\multicolumn{3}{c}
    {\framebox{$3_c 2_L 1_Y$}}\\
$M_Z$ & $\Huge \Downarrow$ & $\langle S\rangle$\\
\multicolumn{3}{c}
    {\framebox{$3_c 1_Q$}}\\
\phantom{$\langle 0_{(6)},1_{(3)}, -1_{(3)}, 0_{(3)}\rangle$ space} & &
\phantom{$\langle 0_{(6)},1_{(3)}, -1_{(3)}, 0_{(3)}\rangle$ space} \\
\end{tabular}
\end{center}
\caption[]{SU(15) symmetry breaking where baryon number is broken by
the VEV of an antisymmetric rank-3 multiplet.   If one considers the
adjoint Higgs multiplet $T$ as a traceless matrix, its VEVs are
diagonal and the notation $1_{(6)}$, e.g.,  stands for six consecutive
entries of unity. In the multiplet $\Phi$, the symbol  $\left<
\hat{d}ue \right>$, e.g., stands for the VEV  of the color singlet
combination of the components with one index having the quantum
numbers of $\hat{d}$, another of $u$ and another of $e$.}
\label{f:153ssb}
\end{figure}

			\begin{figure}
\begin{center}
\Large
\begin{tabular} {rcl}
\multicolumn{3}{c}
    {\framebox{$15$}}\\
$M_G$ & $\Huge \Downarrow$ & $\langle \Phi^{\nu_e
               e^-e^+}\rangle$\\
\multicolumn{3}{c}
    {\framebox{$12_q 3_\ell$}}\\
$M_{12}$ & $\Huge \Downarrow$ & $\langle 1_{(6)}, -1_{(6)},
              0_{(3)}\rangle$\\
\multicolumn{3}{c}
    {\framebox{$6_{qL}6_{qR}1_B 3_\ell$}}\\
$M_{6qL}$ & $\Huge \Downarrow$ & $\langle H^{[ij]}_{[kl]}\rangle$\\
\multicolumn{3}{c}
    {\framebox{$3_{qL}2_{qL}6_{qR}1_B 3_\ell$}}\\
$M_{6qR}$ & $\Huge \Downarrow$
                 & $\langle 0_{(6)},1_{(3)}, -1_{(3)}, 0_{(3)}\rangle$\\
\multicolumn{3}{c}
    {\framebox{$3_{qL}2_{qL}3_{uR}3_{dR}1_B 1_{qR\Lambda} 3_\ell$}}\\
$M_{3\ell}$ & $\Huge \Downarrow$ & $\langle 0_{(12)},1_{(2)}, -2\rangle$\\
\multicolumn{3}{c}
    {\framebox{$3_{qL}2_{qL}3_{uR}3_{dR}1_B 1_{qR\Lambda}
                  2_{\ell L} 1_{\ell Y}$}}\\
$M_B$ & $\Huge \Downarrow$ & $\langle \Delta^{\widehat u\widehat
u\widehat de^+} \rangle$\\
\multicolumn{3}{c}
    {\framebox{$3_{qL} 3_{qR} 2_{qL} 2_{\ell L} 1_F 1_Y$}}\\
$M_S$ & $\Huge \Downarrow$ & $\langle \Phi^{\widehat d
(ue-d\nu_e)}\rangle$\\
\multicolumn{3}{c}
    {\framebox{$3_c 2_L 1_Y$}}\\
$M_Z$ & $\Huge \Downarrow$ & $\langle S\rangle$\\
\multicolumn{3}{c}
    {\framebox{$3_c 1_Q$}}\\
\phantom{$\langle 0_{(6)},1_{(3)}, -1_{(3)}, 0_{(3)}\rangle$ space} & &
\phantom{$\langle 0_{(6)},1_{(3)}, -1_{(3)}, 0_{(3)}\rangle$ space} \\
\end{tabular}
\end{center}
\caption[]{SU(15) symmetry breaking where baryon number is broken by
the VEV of an antisymmetric rank-4 multiplet. The notation for
VEVs has been explained in Fig.~\ref{f:153ssb}.}
\label{f:154ssb}
			\end{figure}

			\begin{figure}
\large
\begin{center}
\begin{picture}(100,65)(-10,20)
\thicklines
\put(0,65){\line(1,-1){15}}     \put(7,58){\vector(1,-1){2}}
\put(6,65){\makebox(0,0){$u$}}		
\put(0,35){\line(1,1){15}}      \put(7,42){\vector(-1,-1){2}}
\put(5,35){\makebox(0,0){$\widehat u$}}	
\multiput(16,50)(4,0){5}{\oval(2,2)[t]}
\multiput(18,50)(4,0){5}{\oval(2,2)[b]}
\put(28,53){\vector(-1,0){4}}
\put(26,45){\makebox(0,0){${\cal G}^{\widehat u}_u$}}
\multiput(35,51)(0,4){5}{\oval(2,2)[r]}
\multiput(35,53)(0,4){5}{\oval(2,2)[l]}
\put(38,58){\vector(0,1){4}}
\put(33,60){\makebox(0,0)[r]{${\cal G}^{\widehat u}_{e^-}$}}
\put(35,70){\line(1,1){15}}	\put(42,77){\vector(1,1){2}}
\put(35,70){\line(-1,1){15}}	\put(27,78){\vector(1,-1){2}}
\put(45,82){\makebox(0,0)[r]{$\widehat u$}}	
\put(25,82){\makebox(0,0)[l]{$e^-$}}		
\multiput(35,50)(3,0){11}{\line(1,0){2}}
\put(45,50){\vector(1,0){2}}
\put(45,48){\makebox(0,0)[t]{$\widehat due^-e^+$}}
\put(62,50){\vector(-1,0){2}}
\put(62,48){\makebox(0,0)[t]{$\nu_ee^-$}}
\put(67,50){\line(1,1){15}}   \put(75,58){\vector(-1,-1){2}}
\put(80,65){\makebox(0,0)[r]{$\nu_e$}}	
\put(67,50){\line(1,-1){15}}    \put(75,42){\vector(-1,1){2}}
\put(80,35){\makebox(0,0)[r]{$e^-$}}	
\multiput(35,50)(0,-3){7}{\line(0,-1){2}}
\put(35,39){\vector(0,1){2}}
\put(35,30){\circle*{3}}
\put(33,30){\makebox(0,0)[r]{$\widehat u\widehat u\widehat de^+$}}
\multiput(55,35)(0,3){10}{\line(0,1){2}}
\multiput(55,35)(0,30){2}{\circle*{3}}		
\put(55,40){\vector(0,-1){2}}
\put(55,32){\makebox(0,0)[t]{$\nu_ee^-e^+$}}
\put(55,60){\vector(0,1){2}}
\put(55,68){\makebox(0,0)[b]{$\widehat due^-$}}
\end{picture}
\end{center}
\caption[]{Tree level diagram giving rise to the operator $\cal O$
of Eq.\  (\ref{154.O1}). The notation about indices has been explained
in Fig.~\ref{f:153O1}.  One can similarly contemplate a
diagram where the VEV of the component $\Phi^{\widehat dd\nu_e}$
appears instead of $\Phi^{\widehat due}$.}\label{f:154pdk}
			\end{figure}
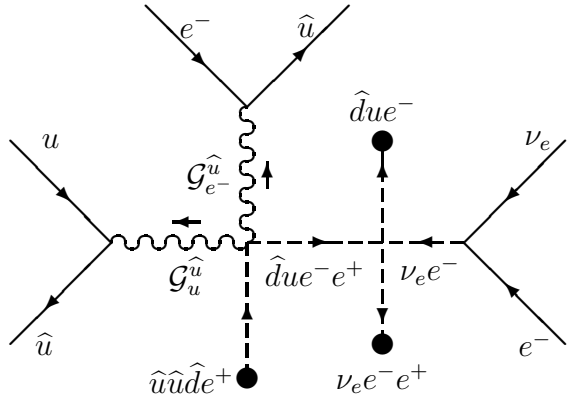

			\begin{figure}
\begin{center}
\Large
\begin{tabular} {rcl}
\multicolumn{3}{c}
    {\framebox{$15$}}\\
$M_G$ & $\Huge \Downarrow$ & $\langle \Phi^{\nu_e
               e^-e^+}\rangle$\\
\multicolumn{3}{c}
    {\framebox{$12_q 3_\ell$}}\\
$M_{12}$ & $\Huge \Downarrow$ & $\langle 1_{(6)}, -1_{(6)},
              0_{(3)}\rangle$\\
\multicolumn{3}{c}
    {\framebox{$6_{qL}6_{qR}1_B 3_\ell$}}\\
$M_{6qL}$ & $\Huge \Downarrow$ & $\langle H^{[ij]}_{[kl]}\rangle$\\
\multicolumn{3}{c}
    {\framebox{$3_{qL}2_{qL}6_{qR}1_B 3_\ell$}}\\
$M_B$ & $\Huge \Downarrow$ & $\langle J^{\widehat d\widehat
d\widehat d\nu_ee^-} \rangle$\\
\multicolumn{3}{c}
     {\framebox{$3_{qL} 2_{qL} 3_{uR} 3_{dR} 2_{\ell L} 1_Y$}}\\
$M_R$ & $\Huge \Downarrow$ & $\langle H^{[ij]}_{[kl]} \rangle$\\
\multicolumn{3}{c}
     {\framebox{$3_{qL} 2_{qL} 3_{qR} 2_{\ell L} 1_Y$}}\\
$M_S$ & $\Huge \Downarrow$ & $\langle \Phi^{\widehat d
(ue-d\nu_e)}\rangle$\\
\multicolumn{3}{c}
    {\framebox{$3_c 2_L 1_Y$}}\\
$M_Z$ & $\Huge \Downarrow$ & $\langle S\rangle$\\
\multicolumn{3}{c}
    {\framebox{$3_c 1_Q$}}\\
\phantom{$\langle 0_{(6)},1_{(3)}, -1_{(3)}, 0_{(3)}\rangle$ space} & &
\phantom{$\langle 0_{(6)},1_{(3)}, -1_{(3)}, 0_{(3)}\rangle$ space} \\
\end{tabular}
\end{center}
\caption[]{SU(15) symmetry breaking where baryon number is broken by
the VEV of an antisymmetric rank-5 multiplet. The notation for
VEVs has been explained in Fig.~\ref{f:153ssb}. The two different VEVs
of the multiplet $H$ have been described in the text.}
\label{f:155ssb}
			\end{figure}


			\begin{figure}
\large
\begin{center}
\begin{picture}(80,40)(-5,30)
\thicklines
\put(0,65){\line(1,-1){15}}     \put(7,58){\vector(1,-1){2}}
\put(6,65){\makebox(0,0)[l]{$\widehat u$}}		
\put(0,35){\line(1,1){15}}      \put(7,42){\vector(1,1){2}}
\put(5,35){\makebox(0,0)[l]{$\widehat d$}}		
\put(70,65){\line(-1,-1){15}}   \put(63,58){\vector(-1,-1){2}}
\put(65,65){\makebox(0,0)[r]{$\widehat u$}}		
\put(70,35){\line(-1,1){15}}    \put(63,42){\vector(-1,1){2}}
\put(65,35){\makebox(0,0)[r]{$e^+$}}		
\multiput(15,50)(3,0){13}{\line(1,0){2}}
\put(25,50){\vector(1,0){2}}
\put(25,52){\makebox(0,0)[b]{$\widehat u\widehat d$}}
\put(45,50){\vector(-1,0){2}}
\put(45,52){\makebox(0,0)[b]{$\widehat ue^+$}}
\multiput(35,50)(0,3){5}{\line(0,1){2}}
\put(35,58){\vector(0,1){2}}
\put(35,65){\circle*{3}}		
\put(35,68){\makebox(0,0)[b]
       {$\widehat u\widehat u\widehat de^+$}}
\end{picture}
\end{center}
\caption[]{Higgs boson mediated diagram giving rise to the
operator of Eq.\ (\ref{15A4.O}).}\label{f:15A4}
			\end{figure}

			\begin{figure}
\large
\begin{center}
\begin{picture}(90,60)(-10,25)
\thicklines
\put(0,65){\line(1,-1){15}}     \put(7,58){\vector(1,-1){2}}
\put(6,65){\makebox(0,0){$d$}}		
\put(0,35){\line(1,1){15}}      \put(7,42){\vector(-1,-1){2}}
\put(5,35){\makebox(0,0){$\widehat d$}}	
\put(70,65){\line(-1,-1){15}}   \put(63,58){\vector(1,1){2}}
\put(65,65){\makebox(0,0){$\widehat d$}}	
\put(70,35){\line(-1,1){15}}    \put(63,42){\vector(-1,1){2}}
\put(65,35){\makebox(0,0){$e^+$}}	
\multiput(16,50)(4,0){10}{\oval(2,2)[t]}
\multiput(18,50)(4,0){10}{\oval(2,2)[b]}
\put(25,52){\makebox(0,0){\vector(-1,0){5}}}
\put(25,47){\makebox(0,0)[t]{${\cal G}^{\widehat d}_d$}}	
\put(45,52){\makebox(0,0){\vector(-1,0){5}}}
\put(45,47){\makebox(0,0)[t]{${\cal G}^{e^+}_{\widehat d}$}}	
\multiput(35,35)(0,3){14}{\line(0,1){2}}
\multiput(35,43)(0,20){2}{\vector(0,1){2}}
\put(37,62){$\widehat d d\nu_ee^-e^+$}		
\multiput(20,76)(3,0){10}{\line(1,0){2}}
\put(26,76){\vector(-1,0){2}}
\put(44,76){\vector(1,0){2}}
\put(35,35){\circle*{3}}		
\put(35,32){\makebox(0,0)[t]
            {$\widehat d\widehat d\widehat d\nu_ee^-$}}
\put(50,76){\circle*{3}}		
\put(52,76){\makebox(0,0)[l]{$e^-e^+$}}		
\put(20,76){\circle*{3}}		
\put(18,76){\makebox(0,0)[r]{$\widehat dd\nu_e$}}
\end{picture}
\end{center}
\caption[]{Gauge boson mediated diagram giving rise to the operator
of Eq.\ (\ref{15A5.O1}).}\label{f:15A5g}
			\end{figure}


			\begin{figure}
\large
\begin{center}
\begin{picture}(110,40)(-10,30)
\thicklines
\put(0,65){\line(1,-1){15}}     \put(7,58){\vector(1,-1){2}}
\put(6,65){\makebox(0,0)[l]{$\widehat d$}}		
\put(0,35){\line(1,1){15}}      \put(7,42){\vector(1,1){2}}
\put(5,35){\makebox(0,0)[l]{$\widehat d$}}		
\put(90,65){\line(-1,-1){15}}   \put(83,58){\vector(-1,-1){2}}
\put(85,65){\makebox(0,0)[r]{$\widehat d$}}		
\put(90,35){\line(-1,1){15}}    \put(83,42){\vector(-1,1){2}}
\put(85,35){\makebox(0,0)[r]{$e^-$}}		
\multiput(15,50)(3,0){20}{\line(1,0){2}}
\put(25,50){\vector(1,0){2}}
\put(25,48){\makebox(0,0)[t]{$\widehat d\widehat d$}}
\put(45,50){\vector(-1,0){2}}
\put(45,48){\makebox(0,0)[t]{$\widehat d\nu_ee^-$}}
\put(65,50){\vector(-1,0){2}}
\put(65,48){\makebox(0,0)[t]{$\widehat de^-$}}
\multiput(35,50)(0,3){5}{\line(0,1){2}}
\put(35,58){\vector(0,1){2}}
\multiput(55,35)(0,3){10}{\line(0,1){2}}
\put(55,43){\vector(0,-1){2}}
\put(55,58){\vector(0,-1){2}}
\put(35,65){\circle*{3}}		
\put(35,68){\makebox(0,0)[b]
       {$\widehat d\widehat d\widehat d\nu_ee^-$}}
\multiput(55,35)(0,30){2}{\circle*{3}}		
\put(55,32){\makebox(0,0){$e^-e^+$}}
\put(55,68){\makebox(0,0)[b]{$\nu_ee^-e^+$}}
\end{picture}
\end{center}
\caption[]{Higgs boson mediated diagram giving rise to the
operator of Eq.\ (\ref{15A5.O2}).}\label{f:15A5h}
			\end{figure}

			\begin{figure}
\begin{center}
\Large
\begin{tabular} {rcl}
\multicolumn{3}{c}
    {\framebox{$16$}}\\
$M_G$ & $\quad\Huge \Downarrow\quad$ & $\langle \Delta^{\nu_e
               e^-e^+ \widehat\nu_e}\rangle$\\
\multicolumn{3}{c}
    {\framebox{$12_q 4_\ell$}}\\
$M_{12}$ & $\Huge \Downarrow$ & $\langle 1_{(6)}, -1_{(6)},
              0_{(4)}\rangle$\\
\multicolumn{3}{c}
    {\framebox{$6_{qL}6_{qR}1_B 4_\ell$}}\\
$M_{6qL}$ & $\Huge \Downarrow$ & $\langle H^{[ij]}_{[kl]}\rangle$\\
\multicolumn{3}{c}
    {\framebox{$3_{qL}2_{qL}6_{qR}1_B 4_\ell$}}\\
$M_{6qR}$ & $\Huge \Downarrow$
                 & $\langle 0_{(6)},1_{(3)}, -1_{(3)}, 0_{(4)}\rangle$\\
\multicolumn{3}{c}
    {\framebox{$3_{qL}2_{qL}3_{uR}3_{dR}1_B 1_{qR\Lambda} 4_\ell$}}\\
$M_B$ & $\Huge \Downarrow$ & $\langle \Delta^{\widehat u\widehat
d\widehat d\widehat\nu} \rangle$\\
\multicolumn{3}{c}
    {\framebox{$3_{qL}2_{qL}3_{qR}1_{qY} 3_\ell$}}\\
$M_{3\ell}$ & $\Huge \Downarrow$ & $\langle 0_{(12)},1_{(2)}, -2,0\rangle$\\
\multicolumn{3}{c}
    {\framebox{$3_{qL}2_{qL}3_{qR}1_{qY}
                  2_{\ell L} 1_{\ell Y}$}}\\
$M_S$ & $\Huge \Downarrow$ & $\langle \Delta^{\widehat d \widehat\nu
(ue - d\nu_e)}\rangle$\\
\multicolumn{3}{c}
    {\framebox{$3_c 2_L 1_Y$}}\\
$M_Z$ & $\Huge \Downarrow$ & $\langle S\rangle$\\
\multicolumn{3}{c}
    {\framebox{$3_c 1_Q$}}\\
\phantom{$\langle 0_{(6)},1_{(3)}, -1_{(3)}, 0_{(3)}\rangle$ space} & &
\phantom{$\langle 0_{(6)},1_{(3)}, -1_{(3)}, 0_{(3)}\rangle$ space} \\
\end{tabular}
\end{center}
\caption[]{SU(16) symmetry breaking where baryon number is broken by
the VEV of an antisymmetric rank-4 multiplet. The notation for
VEVs has been explained in Fig.~\ref{f:153ssb}.}\label{f:164ssb}
	\end{figure}

			\begin{figure}
\begin{center}
\Large
\begin{tabular} {rcl}
\multicolumn{3}{c}
    {\framebox{$16$}}\\
$M_G$ & $\quad\Huge \Downarrow\quad$ & $\langle \Delta^{\nu_e
               e^-e^+ \widehat\nu_e}\rangle$\\
\multicolumn{3}{c}
    {\framebox{$12_q 4_\ell$}}\\
$M_{12}$ & $\Huge \Downarrow$ & $\langle 1_{(6)}, -1_{(6)},
              0_{(4)}\rangle$\\
\multicolumn{3}{c}
    {\framebox{$6_{qL}6_{qR}1_B 4_\ell$}}\\
$M_{6qL}$ & $\Huge \Downarrow$ & $\langle H^{[ij]}_{[kl]}\rangle$\\
\multicolumn{3}{c}
    {\framebox{$3_{qL}2_{qL}6_{qR}1_B 4_\ell$}}\\
$M_{6qR}$ & $\Huge \Downarrow$
                 & $\langle 0_{(6)},1_{(3)}, -1_{(3)}, 0_{(4)}\rangle$\\
\multicolumn{3}{c}
    {\framebox{$3_{qL}2_{qL}3_{uR}3_{dR}1_B 1_{qR\Lambda} 4_\ell$}}\\
$M_{4\ell}$ & $\Huge \Downarrow$
                 & $\langle \Phi^{\nu_ee^-e^+}\rangle$\\
\multicolumn{3}{c}
    {\framebox{$3_{qL}2_{qL}3_{uR}3_{dR}1_B 1_{qR\Lambda} 3_\ell$}}\\
$M_B$ & $\Huge \Downarrow$ & $\langle \Phi^{\widehat u\widehat
d\widehat d} \rangle$\\
\multicolumn{3}{c}
    {\framebox{$3_{qL}2_{qL}3_{qR}1_{qY} 3_\ell$}}\\
$M_{3\ell}$ & $\Huge \Downarrow$ & $\langle 0_{(12)},1_{(2)},
-2,0\rangle$\\
\multicolumn{3}{c}
    {\framebox{$3_{qL}2_{qL}3_{qR}1_{qY}
                  2_{\ell L} 1_{\ell Y}$}}\\
$M_S$ & $\Huge \Downarrow$ & $\langle \Phi^{\widehat d
(ue - d\nu_e)}\rangle$\\
\multicolumn{3}{c}
    {\framebox{$3_c 2_L 1_Y$}}\\
$M_Z$ & $\Huge \Downarrow$ & $\langle S\rangle$\\
\multicolumn{3}{c}
    {\framebox{$3_c 1_Q$}}\\
\phantom{$\langle 0_{(6)},1_{(3)}, -1_{(3)}, 0_{(3)}\rangle$ space} & &
\phantom{$\langle 0_{(6)},1_{(3)}, -1_{(3)}, 0_{(3)}\rangle$ space} \\
\end{tabular}
\end{center}
\caption[]{SU(16) symmetry breaking where baryon number is broken by
the VEV of an antisymmetric rank-3 multiplet. The notation for
VEVs has been explained in Fig.~\ref{f:153ssb}.}\label{f:163ssb}
	\end{figure}

\end{document}